\documentstyle[aps,12pt,epsf]{revtex}
\newcommand{\lettersize}{\baselineskip=0.5cm}
\newcommand{\dir}{FIGS2}
\newcommand{\fig}[3]
{
\begin{center}
     \noindent
     \unitlength=1mm
     \begin{picture}(#2,#3)
     \put(0,0){\leavevmode \epsfxsize=#2mm \epsffile{\dir/#1}}
     \end{picture}
   \noindent
\end{center}
}
\begin{document}
\baselineskip=12pt
\setcounter{page}{1}
%

\baselineskip=1cm

\noindent
{\LARGE\bf
Phase behavior of grafted chain molecules: \\
Influence of head size and chain length
}

\vspace{0.5cm}

\lettersize

\begin{center}
C. Stadler, F. Schmid \\
{\em Institut f\"ur Physik, Universit\"at Mainz, D-55099 Mainz, Germany}
\end{center}


\begin{quote}
{\bf Abstract.}
Constant pressure Monte Carlo simulations of a coarse grained off-lattice 
model for monolayers of amphiphilic molecules at the air/water interface are 
presented. Our study focusses on phase transitions within a monolayer rather
than on self aggregation. We thus model the molecules as stiff chains
of Lennard-Jones spheres with one slightly larger repulsive end bead (head) 
grafted to a planar surface. Depending on the size of the head, the 
temperature and the pressure, we find a variety of phases, which differ in 
tilt order (including tilt direction), and in positional order. 
In particular, we observe a modulated phase with a striped superstructure. 
The modulation results from the competition between two length scales, the 
head size and the tail diameter. As this mechanism is fairly general, it
may conceivably also be relevant in experimental monolayers. We argue that 
the superstructure would be very difficult to detect in a scattering 
experiment, which perhaps accounts for the fact that it has not been reported 
so far. Finally the effect of varying the chain length on the phase
diagram is discussed. Except at high pressures and temperatures, the
phase boundaries in systems with longer chains are shifted to higher
temperatures.

\end{quote}

\lettersize

\section{Introduction}

Amphiphilic molecules on water often form monomolecular layers (Langmuir 
monolayers): The nonpolar tails prevent them from dissolving in the water, and
the water soluble head groups cause them to spread on the water instead
of agglomerating into droplets. Such monolayers are of interest for
various reasons: As model systems which can teach about the structure
of lipid bilayers, omnipresent in every living organism; 
as model systems which display all kinds of ordering phenomena in 
two dimensions. Amphiphile monolayers on solid substrates 
(Langmuir-Blodgett films\cite{lb-books}) are studied because of their
possible use in electronic and opto-electronic devices.

If the tails are sufficiently long, Langmuir monolayers show a rich spectrum 
of phases\cite{kaganer-review,mono-reviews}, depending on the temperature 
and the area density of the molecules. Among these are a disordered fluid 
state (liquid expanded) and numerous condensed phases with varying degree of 
intrinsic order. For example, the low-temperature phases are crystalline, 
and those at higher temperatures are hexatic (``liquid condensed''):
The positional order is lost, but the orientations of the bonds connecting
nearest neighbors have long-range correlations\cite{halperin-nelson}. The 
chains in the low-pressure phases are often collectively tilted in one 
direction, whereas they stand on average upright in the high-pressure phases.
Other types of order exist as well. The characteristics of the phase diagrams 
are in many respect similar for lipids, fatty acids, esters and alcohols. 
Experimental results suggest some general rules: 
Adding CH${}_2$ units to the hydrocarbon chains has the same effect as reducing 
the temperature. Playing with the size of the head group ({\em e.g}, by 
varying the pH of the subphase\cite{shih1}, or by exchanging the head 
group\cite{shih2,teer}) basically affects the tilting transitions. Moreover, 
there seems to be a relation between the effective size of the head groups and 
the direction of tilt\cite{teer}. We will come back to that point shortly.

These general features are common to monolayers of very different 
molecules, which share only a few distinctive properties, such as the 
existence of a polar head and of at least one flexible nonpolar tail. 
Hence studies of idealized models seem to be a promising approach
to explore the connection between the monolayer properties and the
molecular structure. Popular coarse-grained models for amphiphilic 
molecules are, for example, Lennard-Jones chains\cite{smit1,smit2}:
Beads of two different types are connected together with springs to 
form the nonpolar ``tail'' units and the polar ``head units'' units. 
Bead-spring models have been employed to study 
self-assembling micelles\cite{smit2,palmer,gottberg,viduna,goetz}
properties of bilayers\cite{goetz,benshaul} 
and monolayers\cite{haas2,haas3,christoph0,christoph-diss,christoph1}.
If one is interested in the internal structure rather than in self-assembly,
the latter are most simply modelled by chains attached to one head unit, which
is confined to a surface. A monolayer model of this kind has first been 
investigated in simulations by Haas, Hilfer and Binder\cite{haas2,haas3} 
and later by us\cite{christoph0,christoph-diss,christoph1}. 
A related model of entirely stiff chains has been studied by 
Opps and coworkers\cite{opps}.

Collective tilt can be induced in a way which is closest to nature by making 
the heads slightly larger than the tail diameter. Even if heads and tail 
monomers are identical, chains sometimes tilt because of intermolecular 
packing effects -- tilted chains lock better into each other. However, this 
tilting mechanism depends strongly on the internal chain structure and thus 
on the particular choice of the model parameters. The relation between
chain stretching, chain packing, and chain tilting are far 
from transparent\cite{haas3}. If the tilt is induced by a mismatch of head and 
tail size, on the other hand, many properties of the tilted state can be 
understood from simple geometrical arguments. For example, the relation between
the effective head group size, the surface pressure, and the direction of tilt 
mentioned earlier can be rationalized as follows: 
If the heads are just slightly larger than the tails, the chains can optimize 
the distance to four neighbor chains in the state tilted towards next nearest 
neighbors, but only to two neighbor chains in the state tilted towards nearest 
neighbors. Tilt towards next nearest neighbors is thus more favorable.
When the head group size reaches a certain value, the chains turn to tilt
towards nearest neighbors. This is because the hexagonal lattice of the chains 
gets distorted in the direction of tilt. In the next nearest neighbor tilted 
state, two distances remain unaffected by the distortion. In the nearest 
neighbor tilted state, however, all six distances to nearest neighbors increase
as a result, and the larger head groups can settle in happily in the larger 
cage. The argument can be cast into a simple
mathematical formula\cite{metilt}. It predicts that chains with sufficiently 
large heads tilt towards nearest neighbors at low surface pressures, and
towards next nearest neighbors at higher pressures. This is exactly what
is found experimentally\cite{kaganer-review}.

In a recent paper\cite{christoph1}, we have presented extensive Monte
Carlo simulations of a bead-spring model with a 10 \% larger head bead
attached to a planar surface. We have calculated the phase diagram in
pressure-temperature space and in the space of molecular area vs. temperature.
The model exhibited a disordered fluid and a number of condensed phases, an 
untilted one (U), one with tilt towards next nearest neighbors (NNN), and at
low pressures one with tilt towards nearest neighbors (NN). Hence the
model reproduces some of the important states of Langmuir monolayers and
can be considered as a useful minimal model for the study of such systems.
The phase diagram in pressure-temperature space is shown in Fig. \ref{fig1}.  
The goal of the present study was to extend that work in order to gain
a better understanding of the connection between the chain structure
and the phase behavior. To this end, we have varied the size of the
head bead and the chain length. Surprisingly, we found that for heads which
are more than 14 \% larger than the tails, a new phase emerges with a 
modulated superstructure and an average tilt direction which is intermediate 
between nearest and next nearest neighbors. Such a superstructure has not been 
reported experimentally; however, we will argue that it would probably be 
difficult to identify with the available experimental methods. We will
characterize the modulated phase and present a detailed phase diagram
for a system with 20 \% larger head beads. Then we will study the effect on
the phase transition of varying the chain length and comment on finite-size 
effects. Systems with very large heads shall also be discussed very briefly.

\section{The model}

\label{sec2}

We study a system of short chains of length $N$, with one larger head
bead confined to move within a planar surface (The $xy$ plane, $z=0$).
Beads are not allowed to enter the half space $z<0$. 
They interact with a truncated Lennard-Jones potential
\begin{equation}
\label{vlj}
V_{LJ}(r) = \left\{ \begin{array}{lcr}
\epsilon \cdot 
\Big( \big(\sigma/r\big)^{12} - 2 \big(\sigma/r\big)^6 + v_c \Big) &
\mbox{for} & r \le c \; \sigma \\
0 & \mbox{for} & r > c \; \sigma
\end{array} \right. ,
\end{equation}
where the offset $v_c = 2 c^{-6} - c^{-12}$ is chosen such that 
$V_{LJ}(r)$ is continuous at the cutoff $r= c \; \sigma$. 
As long as the cutoff parameter $c>1$, the interactions have an attractive 
part. Here, we use $c=2$ for the tail bead interactions. The head beads are 
made purely repulsive by choosing $c=1$ for the head-head and the head-tail
interactions. Furthermore, the head diameter is slightly larger than the tail 
diameter, $\sigma_H > \sigma$. The effective diameter for the head-tail 
interactions is $(\sigma_H + \sigma)/2$. Beads in a chain are connected
by springs of equilibrium length $d_0$ with the spring potential
\begin{equation}
\label{fene}
V_{S}(d) = \left\{ \begin{array}{l c r}
- \frac{k_{S}}{2} \; d_{S}^2 \; \ln\Big[ 1 - (d-d_0)^2/d_{S}{}^2 \Big]
& \mbox{for} & |d-d_0|<d_{S} \\
\infty & \mbox{for} & |d-d_0| > d_{S}
\end{array} \right. 
\end{equation}
(FENE potential). The chains are made stiff with a stiffness potential
\begin{equation}
\label{ba}
V_{A} = k_{A} \cdot (1- \cos \theta),
\end{equation}
which acts on the angle $\theta$ between subsequent springs and favors 
$\theta=0$ (straight chains). 

We have performed Monte Carlo simulations of $n$ chains grafted on a 
parallelogram with periodic boundaries in the $xy$ direction, under
conditions of constant spreading pressure $\Pi$  in the $xy$ plane. 
In order to avoid internal shear stress, both side lengths $L_x$ and
$L_y$ and the angle $\alpha$ of the parallelogram were allowed to
fluctuate during the simulations. The Monte Carlo moves include 
single monomer displacements, and rescaling of all coordinates such that 
$L_x$, $L_y$, or $\alpha$ changes. They are accepted or rejected according
to a Metropolis prescription with the effective 
Hamiltonian\cite{allen-tildesley}
\begin{equation}
\label{heff1}
H = E + \Pi A - n N T \ln(A),
\end{equation}
where $E$ is the internal energy, $\Pi$ the applied spreading pressure, and
$A = L_x L_y \sin \alpha$ the area of the simulation box. 
During the simulations, we have monitored the ``internal pressure tensor''
\begin{equation}
{\Pi}^{int}_{\alpha \beta} = \frac{1}{A} 
\langle \sum_{i=1}^{n N} r_{i \alpha} {F}_{i \beta} \rangle
 + \frac{N k_B T}{A} \delta_{\alpha \beta},
\end{equation}
and checked that it is diagonal and identical to $\Pi \delta_{\alpha \beta}$
as it should. Here the sum $i$ runs over all monomers, $\alpha, \beta$ over 
the $x$ and $y$ coordinate, $\vec{F}_i$ denotes the force acting on monomer 
$i$, and $\delta_{\alpha \beta}$ is the unit matrix. 

In addition to the simulations, we have also performed a low temperature
phonon expansion. The zero temperature ground state, which minimizes
the enthalpy (\ref{heff1}), is characterized by the minimum enthalpy $H_0$
and by the ground state configuration 
$\{ \{\vec{r}_i^0\},L_x^0,L_y^0,\alpha^0\}$. Note that it is continuously
degenerate with respect to the simultaneous translation of all monomers 
in the $x$ or $y$ direction (or, alternatively, the translation of the 
of the simulation box). Hence one has $f=3 n N - n + 1$ nontrivial degrees 
of freedom left. At low temperatures, we approximate the enthalpy by the 
harmonic expansion
\begin{equation}
\label{harm}
H_{harm} = H_0 + \frac{1}{2} \; \mbox{\it \bf u $\,$ M $\,$ u},
\end{equation}
\begin{displaymath}
\mbox{where} \qquad
\mbox{\it \bf u} =\Big\{ \{(\vec{r}_i-\vec{r}_i^0)/\sigma\},
(L_x-L_x^0)/\sigma,(L_y-L_y^0)/\sigma,\alpha-\alpha^0\}
\end{displaymath}
is a dimensionless vector of (supposedly small) deviations from the ground
state, and
\begin{displaymath}
M_{ij} = \frac{\partial^2 H}{\partial u_i \partial u_j}
\end{displaymath}
the matrix of second derivatives. It has $f+2$ Eigenvalues $\varepsilon_s$,
two of which are zero due to the translational invariance mentioned 
above. For a ground state with uniform tilt, {\it \bf M} can be diagonalized 
without too much computational effort, since already a simple Fourier transform
in the $xy$ plane partly diagonalizes it. In the case of the modulated phase, 
we preferred to keep all the coordinates explicitly. Having obtained the
Eigenvalues, one can rewrite (\ref{harm}) as
\begin{equation}
H_{harm} = H_0 + \frac{1}{2} \sum_{s=1}^f \varepsilon_s \zeta_s^2,
\end{equation}
where the sum now runs only over the nonzero eigenvalues, and $\zeta_s$
are the corresponding normal coordinates. Using this expression, one can
calculate the partition function
\begin{eqnarray*}
{\cal Z}& =& \frac{1}{n! \; \sigma^{f-1}}
\int_0^{\infty} d L_x \int_0^{\infty} d L_y \int_0^{\pi} d \alpha
\int_{\mbox{\tiny simulation box}}' d \vec{r}_i e^{-\beta H} \\
& \approx & \frac{1}{n!}\; 
\prod_{s=1}^f \sqrt{\frac{\pi}{\beta \varepsilon_s}} \cdot e^{-\beta H_0}.
\end{eqnarray*}
Here $\beta = 1/k_B T$ is the Boltzmann factor, and the prime indicates that 
the trivial contribution of translating the simulation box in the $xy$ plane
has been omitted in the integral. One obtains the free energy
\begin{equation}
\label{gibbs}
\beta G = - \ln {\cal Z}
= \beta H_0  + \frac{1}{2} \sum_{s=1}^f \ln \beta \varepsilon_s
+ \mbox{const.}
\end{equation}
The approximation is justified as long as the enthalpy measured from simulations
agrees with the prediction of the harmonic approximation
\begin{equation}
H(T) \approx H_0 +  \frac{f}{2} k_B T.
\end{equation}
In our simulations, this was the case for $k_B T/\epsilon < 0.1$.
Note that even for temperatures $T \to 0$, the harmonic approximation is not 
applicable in every respect. For example, it yields the wrong thermal
expansion coefficient\cite{ashcroft-mermin}. However, this is no serious 
problem since we we use it mainly to calculate $\beta G$ at some point in 
phase space $(\Pi_0,T_0)$ here. 
Given this reference value, one can determine $G$ at other temperatures and 
pressures from simulations by thermodynamic integration 
\begin{equation}
G(\Pi,T) = G(\Pi_0,T_0) + k_B T \int_{\Gamma} \Big\{
d \Pi' \frac{A}{k_B T'} - d T' \frac{H}{k_B T'^2} \Big\},
\end{equation}
as long as the path $\Gamma$ from $(\Pi_0,T_0)$ to $(\Pi,T)$ does not cross a 
first order phase transition. 

The comparison of the free energies in the different phases allowed to pinpoint 
the discontinuous phase transitions between condensed phases, even though
some of them were highly metastable over large portions of the phase space.
If one transition point is known, lines of first order transitions can also 
be traced with the Clausius Clapeyron equation
\begin{equation}
\label{clausius}
\frac{\delta \Pi}{\delta T} = \frac{\Delta H}{T \Delta A},
\end{equation}
where $\Delta H$ is the enthalpy difference between the coexisting phases,
and $\Delta A$ the area difference. Applying the Clausius Clapeyron equation
is equivalent to performing a thermodynamic integration along a very special
path, the transition line. We have located phase transitions with both methods, 
by use of eqn. (\ref{clausius}), and by thermodynamic integration with
different paths. The agreement was generally very good.

The parameters of the model are chosen as in Ref. \cite{haas2,haas3} and in
our previous work\cite{christoph1}: $d_0 = 0.7 \sigma, d_S = 0.2 \sigma,
k_S = 100 \epsilon$, and $k_A = 10 \epsilon$. Unless stated otherwise, our
results refer to systems of $n=144$ chains with length $N=7$. These systems
decorrelate on average within 200-1000 Monte Carlo steps (MCS), where
one Monte Carlo step consists of $Nn = 1008$ attempted monomer displacements
and one attempt to rescale $L_x$, $L_y$, and $\alpha$. The systems were
usually equilibrated over 70.000 MCS, and data were collected every 
500st configuration over a period of 200.000 MCS or more.

\section{Results}

\subsection{The modulated phase}

The phase diagram for $\sigma_H=1.1$, {\em i.e.}, heads which are 10 \% larger 
than the tail diameter, has been discussed in detail in Ref. \cite{christoph1}.
It includes a disordered phase (LE) and a number of condensed phases with or
without uniform tilt in one direction (LC-U, LC-NN and LC-NNN, see 
Fig. \ref{fig1}). Such phases have been found numerous times in experimental 
monolayers\cite{kaganer-review,mono-reviews}. On increasing the head 
size, we observe at low pressures a novel phase with a modulated, 
striped superstructure. Figure \ref{fig2} shows the projection of such a 
configuration onto the $xy$ plane. The modulation turns out to be very stable: 
Once a system has assumed a modulated state, it stays modulated even if the 
temperature is raised almost to the melting temperature, and up to pressures 
of $\Pi \approx 90 \epsilon/\sigma^2$ are applied, where the equilibrium phase
transition to a uniformly tilted phase is long passed. On the other hand, 
uniformly tilted states are metastable deep in the region of thermodynamic 
stability of the modulated phase. Hence hysteresis effects are strong. 
At chain length $N=7$, the modulated state is observed for head sizes between 
$1.14  \stackrel{<}{\sim} \sigma_H \stackrel{<}{\sim} 1.27$. 
(If the heads are made even larger, the tails do not form closed tilted layers 
any more, but aggregate into small micelle-like structures.)
We have also found it in systems of chains with length $N=6$ and $N=8$. The 
width of the stripes increases weakly with increasing pressure. A reliable 
determination of the equilibrium stripe width is however 
difficult because of the periodicity imposed by the boundary conditions. 
For the special case $\sigma_H = 1.2 \sigma$ and $N=8$, we have varied the 
system size from 64 to 256 chains, and found that the most favored stripe 
width within the region of thermodynamic stability of the modulated phase is 
approximately four.

A closer inspection of Fig. \ref{fig2} gives some insight in the 
origin of the modulation: The chains arrange themselves in rows, and the
heads (crosses) form a hexagonal lattice which is slightly tilted with 
respect to the rows built by the chains. By means of such a construction, the
distance between the rows can be kept smaller than the distance between
the heads, which is favorable from the point of view of chain packing. 
The modulation thus stems from a competition between two 
incommensurable lengths, the head size and the chain diameter.
It is penalized by lines of ``defects'' at the boundaries of the stripes,
where the heads are localized right between two rows. The chains there have
to bend strongly so that at least their last segments can join the rows. 
Note that the transition from a modulated state to a uniformly tilted
state involves a rearrangement of either the head lattice or the
chain rows, which probably explains the strong hysteresis effects. 
From the above discussion, one might suspect that chain flexibility is
needed to stabilize the modulated phase. However, a very similar structure 
has been observed by Opps {\em et al} in systems of entirely stiff 
rods\cite{opps}. Nevertheless, we expect that the chain stiffness will 
significantly affect the region of stability in phase space of the
modulated phase.

Experimentally, superstructures like the modulation are often identified
by satellite peaks in the structure factor, defined as
\begin{equation}
S(\vec{q}) = \frac{1}{n N} 
\bigg| \sum_{j=1}^{n N} \exp(i \vec{q} \vec{r}_j )\bigg|^2,
\end{equation}
where the sum runs over all monomers in the system. We shall briefly
discuss the form of $S(\vec{q})$ in the different ordered
phases\cite{kaganer-review,christoph1}. The structure factor of
an untilted state has peaks in the $xy$ plane at $q_z=0$ which 
correspond to the reciprocal lattice of the hexagonal lattice. These
Bragg rods\cite{footnote} are very sharp in the $x$ and $y$ direction and 
broad in the $z$ direction, with a width which is inversely proportional to 
the layer thickness. In the uniformly tilted state, the Bragg rods move
away from $q_z=0$ on a plane which is perpendicular to the long axis of
the chains\cite{kaganer-review}. In addition, the positions of the rods in the 
$x$ and $y$ direction shift due to the distortion of the lattice. 
In the modulated phase, satellite peaks emerge in the direction 
perpendicular to the stripes. This is demonstrated in Fig. \ref{fig3}.

In grazing incidence x-ray studies of Langmuir monolayers, the situation is 
complicated by the fact that the diffraction patterns average over all domain 
orientations in the monolayer plane (powder average). Hence one only obtains
information on the radial average $S(q_{\parallel},q_z)$, with
$q_{\parallel} = \sqrt{q_x^2 + q_y^2}$. Moreover, the intensity of the
Bragg rods decays rapidly, so that only the lowest order peaks can
be measured. The average structure factor $S(q_{\parallel},q_z)$
in this $q$-range is displayed in Fig. \ref{fig4} (c). For comparison,
Figs. \ref{fig4} (a) and (b) also show the corresponding pictures in
phases with uniform tilt towards nearest and next nearest neighbors. 
In an untilted phase, the six first order Bragg peaks of the 
hexagonal lattice all collapse on one peak centered around $q_z=0$.
In the tilted phases, they split up: Their $q_{\parallel}$ positions
differ due to the distortion of the lattice, and unless they belong to
$\vec{q}_{\parallel}$ vectors perpendicular to the direction of tilt,
they move out in the $q_z$ direction. Note that for symmetry reasons,
some of the peaks still collapse in the states with uniform tilt towards 
nearest and next nearest neighbors. In the modulated state, they are all
separated from each other. However, this would equally be the case in
a state with uniform tilt towards an intermediate direction between
nearest and next nearest neighbors. Moreover, two of the peaks are so close
that they can easily appear as just one peak if the experimental resolution is
not extremely good. The satellite peaks practically vanish at these small
$q_{\parallel}$-vectors. Hence the diffraction pattern of a modulated
phase would most likely be interpreted as one of a phase with uniform
tilt towards nearest neighbors, or as one of a phase with intermediate
tilt direction. Such phases have indeed been reported.

We conclude that if there is no experimental evidence for a modulated
phase, there is also nothing that would exclude such a possibility.
The modulation would be very hard to identify with a diffraction experiment. 
Considering how stable it is in our model over a wide parameter region,
and considering that the underlying mechanism is a very general one -- the
competition between two incommensurable lengths characterizing head group
packing and chain packing -- we believe that there is a good probability that
it might also be present in some real Langmuir monolayers.

\subsection{Phase behavior for chains with head size $\sigma_h=1.2$}

\begin{figure}
\noindent
\fig{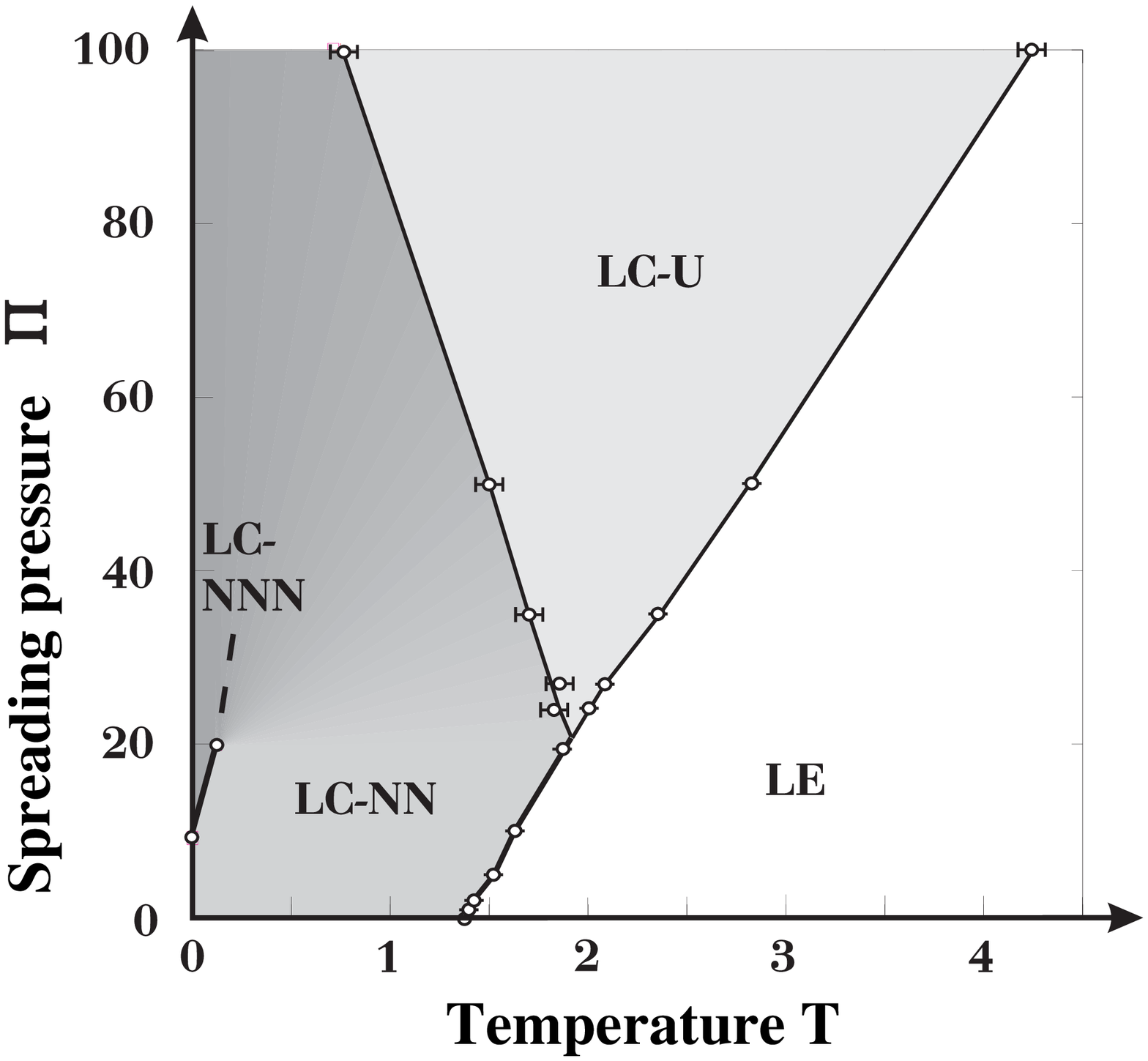}{150}{170} 
\vspace*{-2cm}
\caption{\label{fig1}}
\end{figure}

\begin{figure}
\noindent
\fig{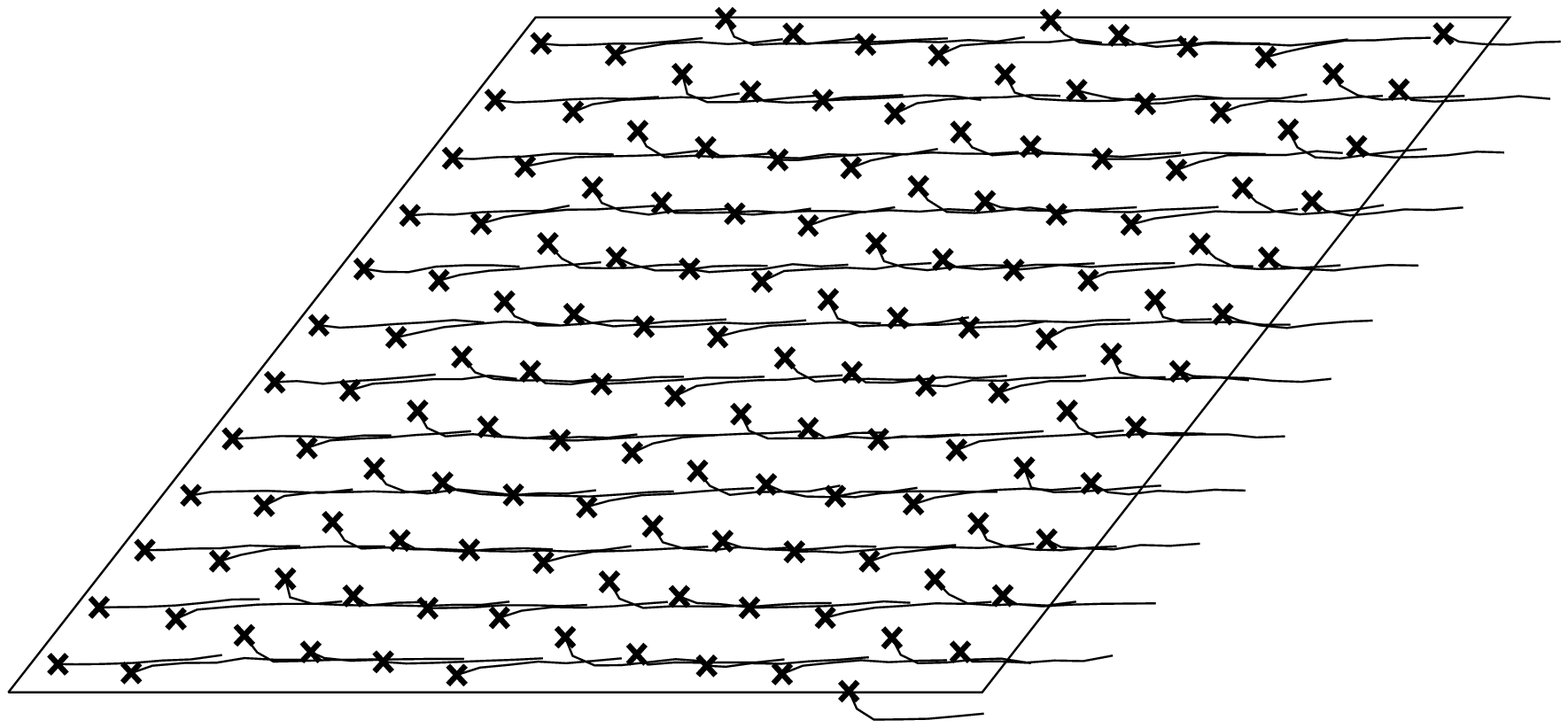}{150}{100} 
\caption{\label{fig2}}
\end{figure}

\begin{figure}
\noindent
\fig{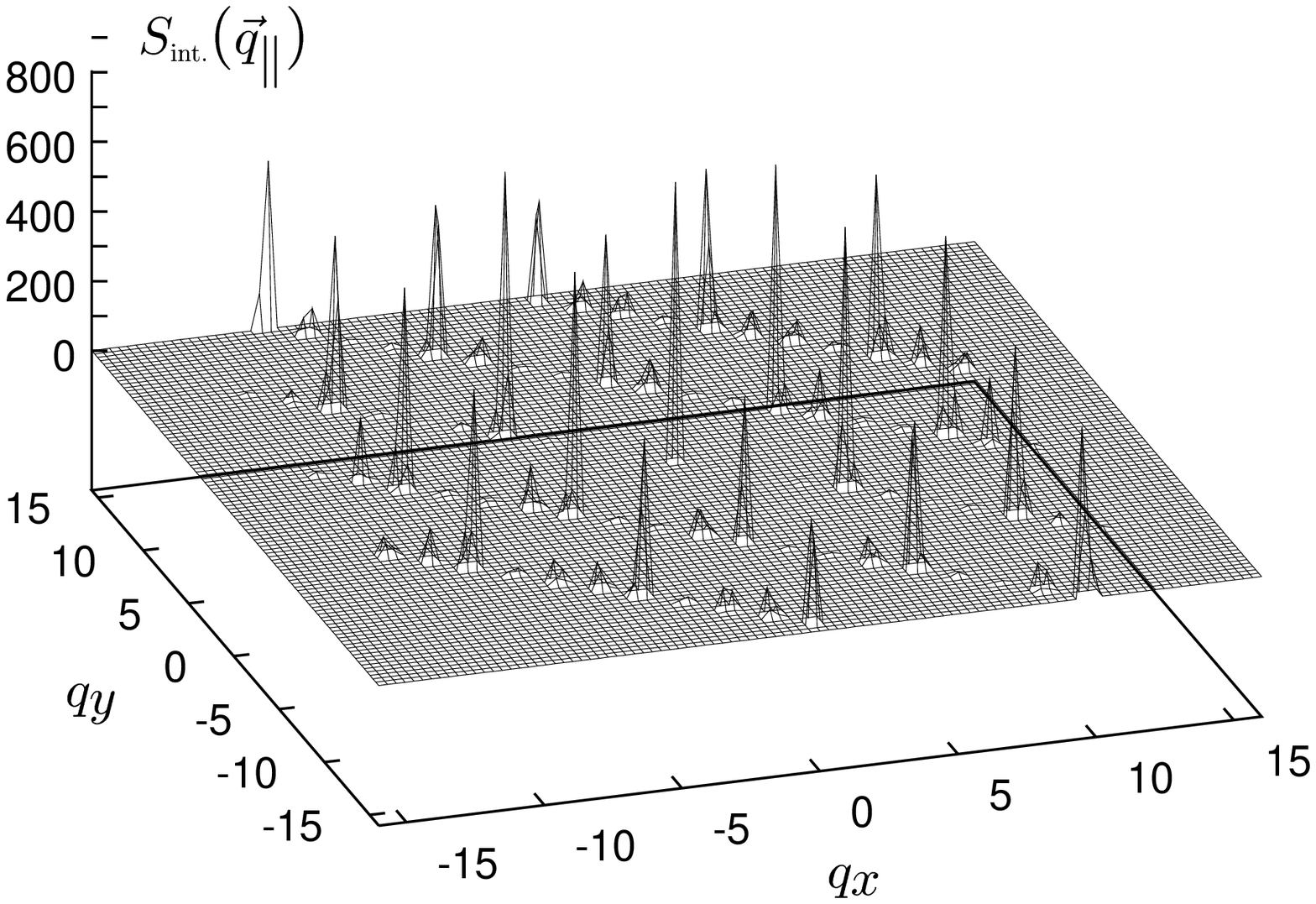}{150}{110} 
\caption{\label{fig3}}
\end{figure}

\begin{figure}
\noindent
(c)\fig{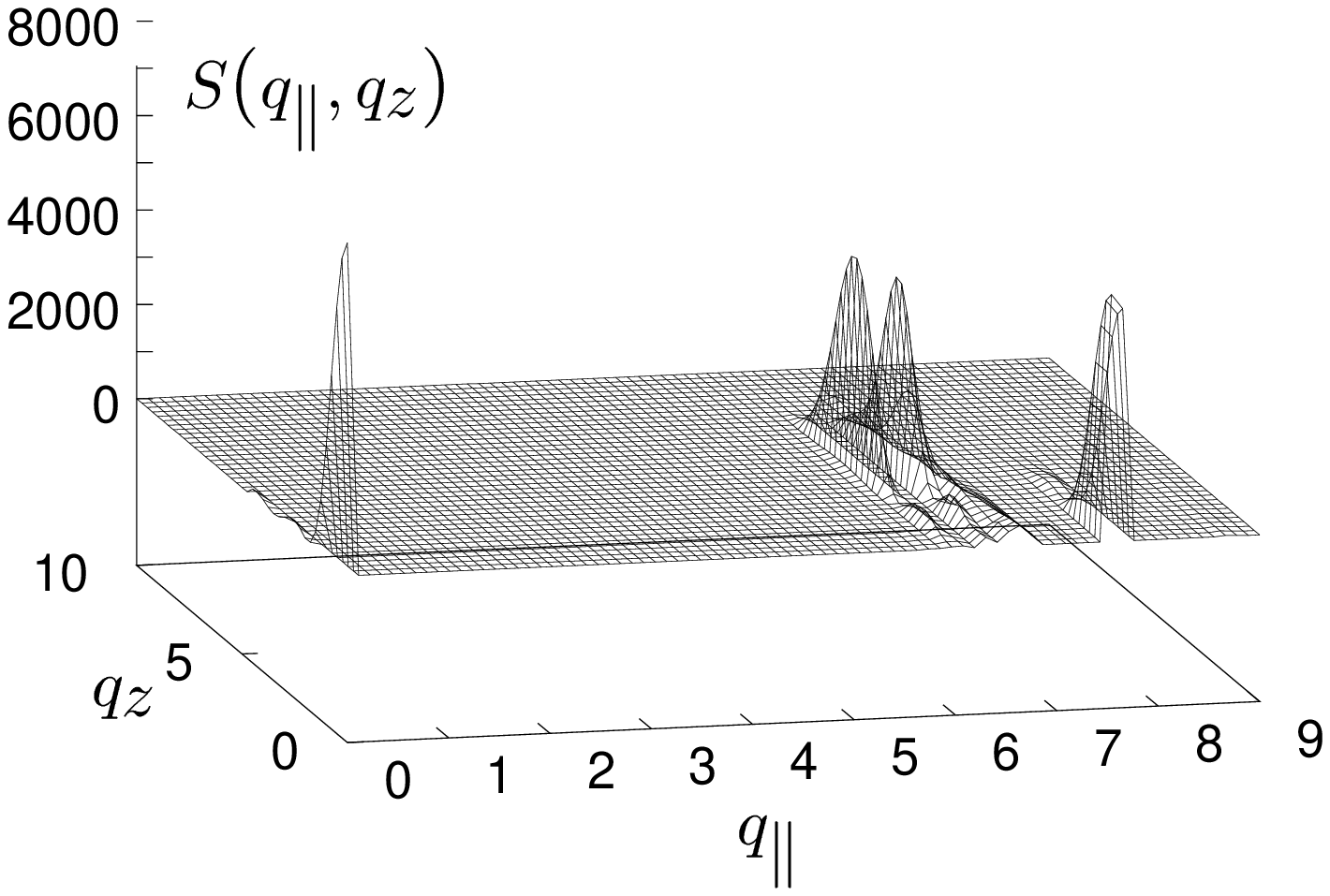}{150}{100} 
\caption{\label{fig4}}
\end{figure}

\begin{figure}
\noindent
\fig{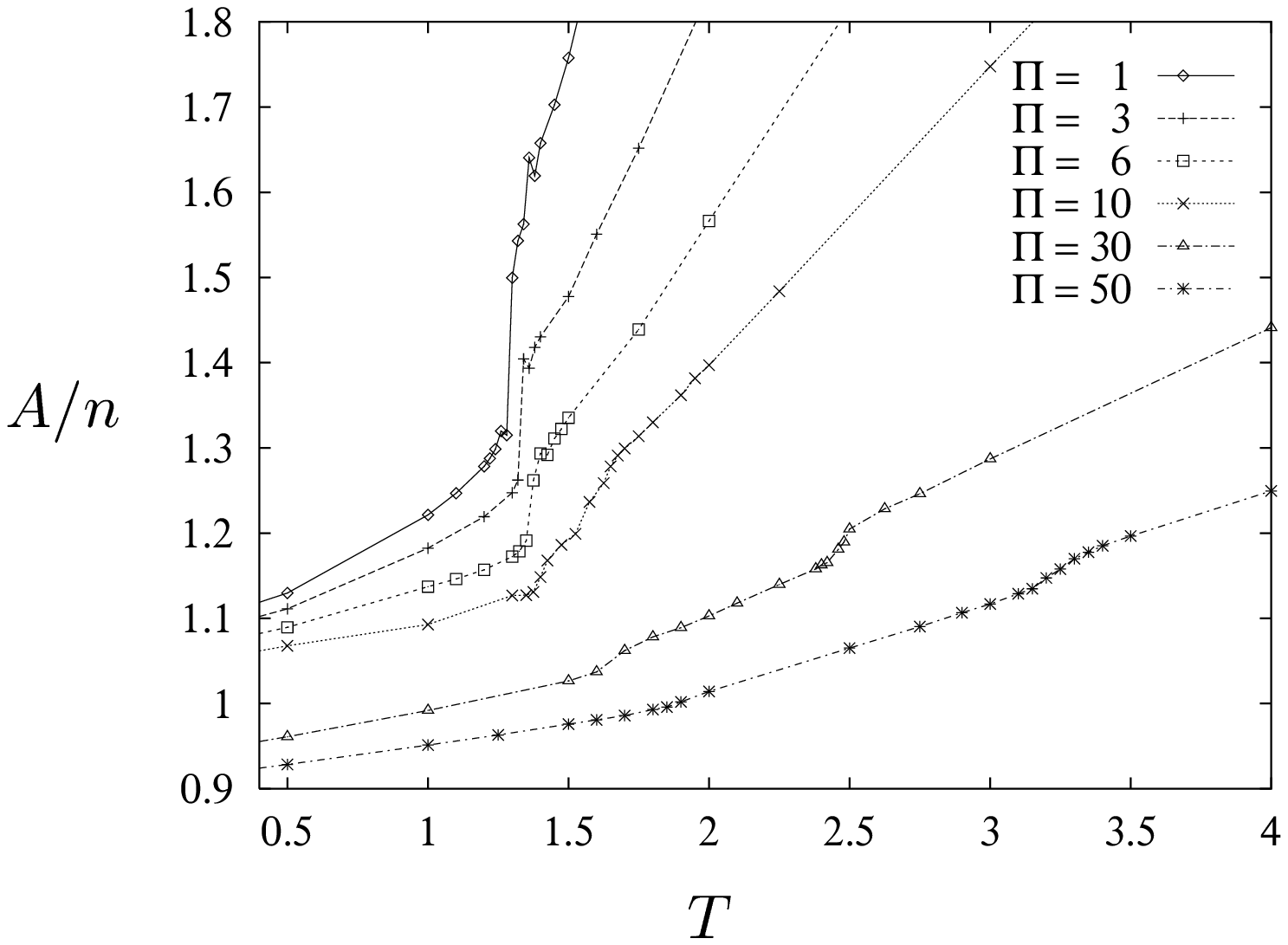}{150}{100} 
\caption{\label{fig5}}
\end{figure}

We have seen that increasing the head size has at least one important effect 
on the phase behavior, it brings the modulated phase into existence.
In order to study further consequences in detail, we have calculated the phase 
diagram for the case of chain length $N=7$ and head size $\sigma_H=1.2$.
Due to the large hysteresis effects mentioned earlier, the phase transition
between the modulated phase and the uniformly tilted state had to be 
determined by thermodynamic integration methods at temperatures 
$T \le 0.8 \epsilon/k_B$. At higher temperatures, hysteresis effects 
disappeared, but the difference between modulation and uniform tilt 
disappeared too, and the transition point became difficult to localize.
For the other phase transitions between tilted and untilted, ordered and
disordered states, hysteresis effects were negligible. These phase transitions
leave their signature in various quantities: Fig. \ref{fig5} shows the
area per molecule as a function of temperature for various pressures.
At low pressures, one identifies one single phase transition accompanied by
a large jump of the molecular area, whereas at high pressures, the 
curves indicate the presence of two singularities. The situation 
thus seems similar to that at head size $\sigma_H=1.1$ (Fig. \ref{fig1}),
where one observes one transition from a tilted ordered to an untilted 
disordered state at low pressures, and two decoupled transitions (one tilting
transition and one order/disorder transition) at high pressures.

\begin{figure}
\noindent
\fig{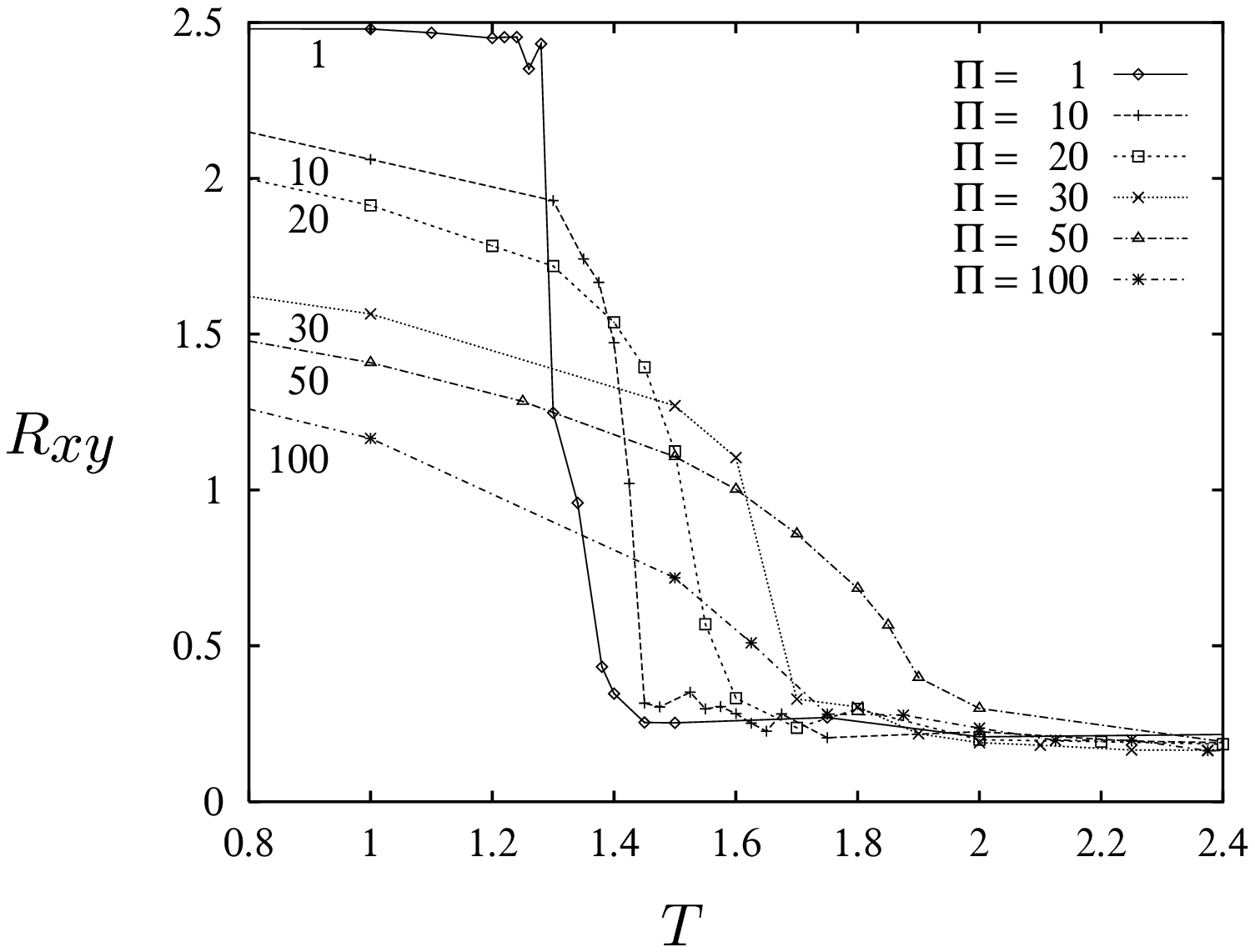}{150}{100}
\caption{\label{fig6}}
\end{figure}

\begin{figure}
\noindent
\fig{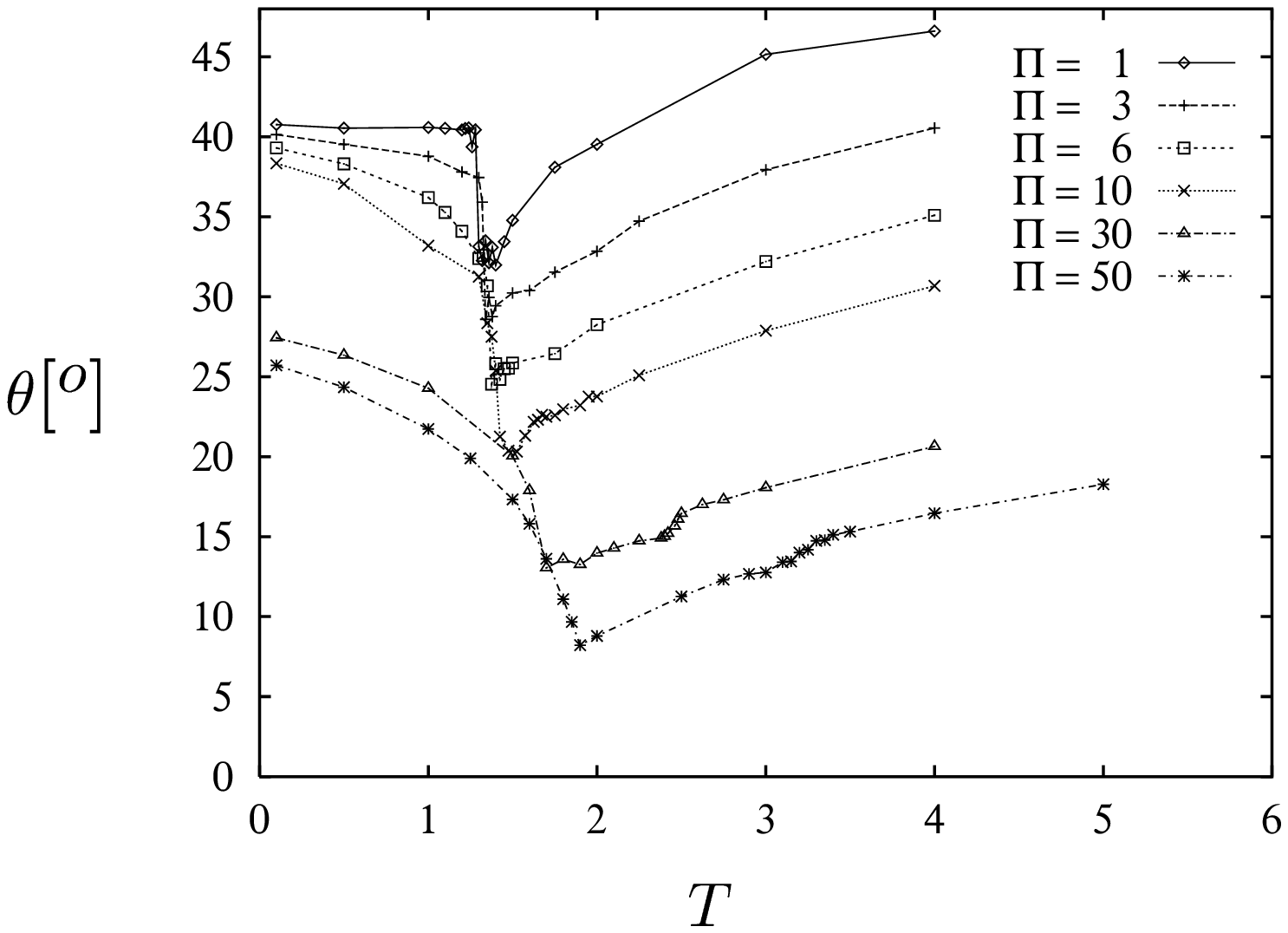}{150}{100} 
\caption{\label{fig7}}
\end{figure}

To check that this interpretation is correct, we have determined
the length of the average projection of the head-to-end vector of the
chains on the $xy$ plane
\begin{equation}
R_{xy} = \sqrt{\langle [x]^2 + [y]^2 \rangle },
\end{equation}
which is a suitable order parameter for collective tilt. Here $[x]$ and $[y]$ 
denote the $x$ and $y$ component of the head-to-end vector, averaged over
the chains of a configuration, and $\langle \cdot \rangle$ denotes
the thermal average over all configurations. It is shown for
various pressures in Fig. \ref{fig6}. As expected, it drops to zero
at the temperature which corresponds to the first singularities in the
area-temperature isobars. One can also inspect average tilt angle $\theta$ 
between the head-to-end vector and the surface normal (Fig. \ref{fig7}). 
Since this quantity does not measure azimuthal symmetry breaking, it is 
nonzero even in the ``untilted'' state. However, it exhibits a sharp
kink at the tilting transition, which is in fact easier to localize than
the kink in $R_{xy}$. (Note that $R_{xy}$ also never reaches strictly zero
due to the finite size of the system.)

\begin{figure}
\noindent
\fig{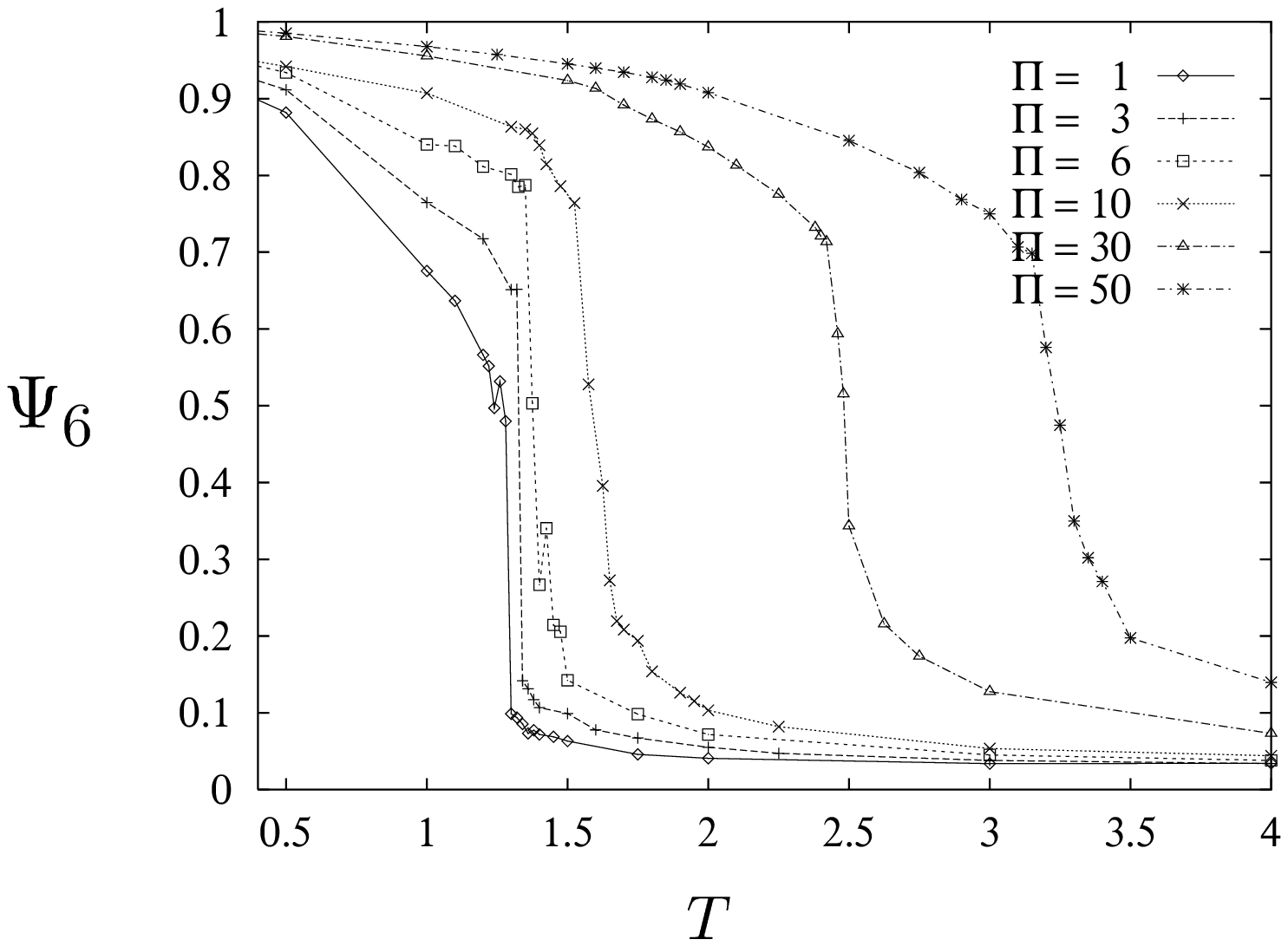}{150}{100} 
\caption{\label{fig8}}
\end{figure}

The quantity which measures the order/disorder transition is the hexagonal
``bond orientational order parameter'' $\Psi_6$ of twodimensional melting,
\begin{equation}
\label{psi}
\Psi_6 = \bigg\langle \bigg| \frac{1}{6n}
\sum_{j=1}^n \sum_{k=1}^6 \exp(i 6 \phi_{jk}) \bigg|^2 \Big\rangle.
\end{equation}
The first sum $j$ runs over all heads of the systems,
the second $k$ over the six nearest neighbors of $j$, and $\phi_{jk}$ is the 
angle between the vector connecting the two heads and an arbitrary 
reference axis. As illustrated in Fig. \ref{fig8}, the parameter $\Psi_6$ 
drops to zero at the temperature corresponding to the second singularity in 
$\theta$ and $A/n$, indicating the transition to a disordered phase where
nearest neighbor directions have no long-range correlations.

The resulting phase diagrams are shown in the pressure-temperature
plane and in the area-temperature plane in Figs. \ref{fig9} and \ref{fig10}.
At low temperatures, the modulated phase supersedes the state with uniform 
tilt towards nearest neighbors. However, the transition pressure to the state 
with uniform tilt towards next nearest neighbor decreases with increasing
temperature, whereas that of the (metastable) transition between two uniformly
tilted states increases. Hence a region is spared out at higher temperatures
where uniform tilt towards nearest neighbors is stable.
Comparing the phase diagram Fig. \ref{fig9} with that of Fig. \ref{fig1}, one 
notices one more qualitative difference: The slope of the transition
from tilted to untilted state is negative for the smaller heads 
$\sigma_H=1.1$, whereas it is partly positive in the system with larger
heads $\sigma_H=1.2$. On increasing the pressure at constant temperature
$T \approx 1.5 \epsilon/k_B$, one observes reentrant behavior: First
a transition from an untilted to a tilted state takes place, then a second
one back to a tilted state. At first sight, it seems strange that the chains 
should start to tilt on being pushed together. Fig. \ref{fig10} suggests a
possible explanation: Tilt is obviously most stable, {\em i.e.}, persists
up to the highest temperatures, at the area per molecule $A/n = 1 \sigma^2$. 
If the chains are squeezed together to lower molecular areas, they stand up. 
On the other hand, if the chains are allowed to occupy more space, they are
in less close contact with each other and disorder more easily. Thus they
remain tilted individually (cf. Fig. \ref{fig7}), but they lose the
collective tilt.

The phase diagram was determined from Monte Carlo simulations of
systems of 144 chains. In order to get a rough estimate of the finite-size
effects, we have performed a few simulations of systems of 900 chains
at pressures $\Pi = 1 \epsilon/\sigma^2$ and $\Pi = 50 \epsilon/\sigma^2$. 
Fig. \ref{fig11} suggests that finite-size effects tend to shift the
apparent melting transition to slightly higher temperatures in small systems.
Yet the shift is so small at pressure $\Pi=50 \epsilon/\sigma^2$ that 
the transition temperatures in the two systems can still be considered the same
within the error. At the lower pressure $\Pi = 1 \epsilon/\sigma^2$, one has 
to keep in mind the additional complication that the ordered state is 
modulated. One would thus expect that some particle 
numbers are more ``commensurate`` with this superstructure than others, and 
systems of different size are not directly comparable. Moreover, the 
relaxation times are very long, and the large systems could not optimize the 
stripe width within the simulation time (400.000 MCS per data point). Hence
it is perhaps not surprising that they were found to melt earlier than the 
smaller system. No systematic finite-size effects were found for the decoupled
tilting transition at pressure $\Pi=50 \epsilon/\sigma^2$ (Fig. \ref{fig12}).

Finite-size effects do of course affect the apparent phase boundaries, and
the actual positions of the boundary lines in Figs. \ref{fig9} and
\ref{fig10} have to be regarded with caution. A determination of the phase 
diagram using systematic finite-size scaling methods would obviously be 
desirable. However, this would not only involve the study of much larger 
systems, but also require a much better statistics of the data.
The computational costs would increase by orders of magnitude. 
We believe that our simulations of a relatively small system and
simulation runs of length 200.000 -- 400.000 MCS are sufficient to
reliably characterize the phase behavior in many respects. Of course,
important questions remain open. For example, we are not able to 
distinguish between a solid state and a hexatic state in the ordered 
region. We have not fully characterized the transition 
from the tilted to the untilted state -- whether it is first order or 
continuous, whether it is one transition or whether one has an intermediate
``unlocked tilted'' state, where the tilt direction is not coupled to the 
local chain lattice etc.  These questions will have to be addressed in future 
studies.

\subsection{Effect of chain length}

Experimentally, it is often observed that increasing the length of
the hydrophobic chains causes the phase transitions to shift to higher
temperatures\cite{kaganer-review}. 
One possible explanation is that the internal energy per molecule scales 
roughly linearly with the chain length, and the entropy gain 
associated with a phase transition is approximately independent of the
chain length. Whereas the first assumption seems reasonable, the second
one is more questionable. It is conceivable for the case of phase
transitions between condensed phases, which affect the orientational degrees 
of freedom of whole chains. However, the melting transition involves
a considerable gain of conformational entropy within the chains, which 
is obviously not chain length independent. 

In order to assess the influence of the chain length on the phase behavior
in our model, we have studied systems of chains with length $N=6, 7$ and
$N=8$ at the spreading pressure $\Pi = 1 \epsilon/\sigma^2$ and
$\Pi = 50 \epsilon/\sigma^2$. The results for the average tilt angle $\theta$
and the order parameter of melting $\Psi_6$ are shown in 
Figs. \ref{fig13} and \ref{fig14}. The transition temperature
of the low-pressure coupled phase transition between the tilted ordered and 
the untilted disordered state at indeed shifts to higher temperatures with 
increasing chain length (Fig. \ref{fig13}). The same holds for the decoupled 
tilting transition at higher pressure (Fig. \ref{fig14} (a)).
The melting transition between the untilted ordered phase and the 
disordered phase, on the other hand, moves to lower temperatures as
the chain length is increased (Fig. \ref{fig14} (b)).

It is instructive to examine the way how the area per molecule depends
on the chain length, Fig. \ref{fig15}. At low pressure,
the area per molecule decreases with increasing chain length at
all temperatures -- presumably an effect of the enhanced attractive
chain interactions (Fig. \ref{fig15} (a)). 
This remains true at $\Pi = 50 \epsilon/\sigma^2$ in the tilted state. 
However, the situation turns around at the tilting transition, and the 
largest chains occupy the largest area at all higher temperatures.
Thus the area per molecule at this high pressure is determined by
the entropic repulsion between the chains rather than by the energetic
attraction between the beads (Fig. \ref{fig15} (b)). 

\section{Summary}

We have examined the phase behavior of a system of stiff short Lennard-Jones
chains grafted on a two dimensional plane with variable head size.
As a function of temperature and surface pressure, we found a disordered
phase and a number of condensed ordered phases with or without
collective chain tilt. The tilting transition and the melting transition
are coupled at low surface pressure and decouple at higher pressures.
In addition, we observe a new modulated phase for a range of head sizes, 
resulting from the competition between the head and the tail size. As we
have argued, there is a possibility that such a phase is also present in real 
monolayers. It would be characterized by a tilt in a direction intermediate
between nearest and next nearest neighbors (which is observed experimentally)
and by satellite peaks in the structure factor at high $q$ vectors
(which are hard to resolve in scattering experiments). With increasing
chain length, most phase transitions are shifted to higher temperatures,
except at high surface pressure the melting temperature. 

The model is versatile enough to provide a useful starting point for
further investigations. Simulations of much larger systems would allow
to study lattice defects and defect interactions, and to distinguish
between a crystalline and a hexatic phase ({\em i.e.}, one with
long range bond orientational order, but only short range positional
correlations\cite{halperin-nelson}). This is of interest because Langmuir 
monolayers are among the few two-dimensional systems where there is strong 
experimental evidence for the existence of hexatic states\cite{helm}. 
Since amphiphile layers often contain amphiphiles of different type, it would 
also be interesting to study mixtures of amphiphiles, {\em e.g.}, long and 
short amphiphiles\cite{chowdhury}, or to add a few amphiphiles with chain 
``defects'', {\em i.e.}, built-in kinks in the chains\cite{levine}. 
Other promising problems relate to the interactions between chains and large
inclusions, {\em e,g}, model ``proteins''\cite{baumgartner1,baumgartner2}, 
and the way how these are affected by the vicinity of phase transitions. 
Furthermore, future studies will be devoted to a more refined treatment
of the head groups. In particular, it will be neccessary to somewhat relax
the rigid constraint of confining them in an entirely flat surface --
by allowing them to move also in the $z$ direction and/or by allowing
for undulations in the substrate.

\section*{Acknowledgments}

We have benefitted from fruitful interactions with Harald Lange and
Kurt Binder. Frank Martin Haas and Rudolf Hilfer let us have their simulation 
code, which has been the base for the development of the code used here.
C.S. thanks the Deutsche Forschungsgemeinschaft (DFG) for support through 
the Graduiertenkolleg on supramolecular systems in Mainz, and F.S. thanks
the DFG for a Heisenberg fellowship.

\clearpage

\noindent
Phase diagram of a model of endgrafted bead-spring chains with 10 \%
larger heads in the pressure-temperature plane. The model is described
in section \ref{sec2}. The spreading pressure $\Pi$ is given in units of 
$\epsilon/\sigma^2$, and the temperature $T$ in units of $\epsilon/k_B$.
LE denotes disordered phase, LC-NN ordered phase with tilt towards nearest 
neighbors, LC-NNN ordered phase with tilt towards next nearest neighbors, 
and LC-U untilted ordered phase. The transition between LC-NN and LC-NNN 
could not be located at pressures above $\Pi = 20 \epsilon/\sigma^2$. 
See text for more explanation.  From Ref. \cite{christoph1}.

\clearpage

\noindent
Projection on the $xy$ plane of a configuration in a modulated state.
Crosses mark the positions of head beads. Parameters are
$\Pi = 1 \epsilon/\sigma^2$, $T=0.1 \epsilon/k_B$, $\sigma_H=1.2 \sigma$.

\clearpage

\noindent
Integrated structure factor $S_{\mbox{\tiny int.}}(\vec{q}_{\parallel}) = 
\int_{q_z^0-\Delta q_z/2}^{q_z^0+\Delta q_z/2} \!  d q_z \; S (\vec{q})$ 
in the modulated phase. For the clarity of the presentation, we have 
integrated over the Bragg rods in the (tilted) Bragg plane
$q_z^0 = - \vec{q}_{\parallel} \vec{e}_t / \cos \theta$, where $\theta$
is the tilt angle, the unit vector $\vec{e}_t$ points towards the
average tilt direction, and $\vec{q}_{\parallel} = (q_x,q_y,0)$.
The interval of integration was $\Delta q_z = 2 \pi/d_0 \cos \theta$.
One clearly discernes rows of satellite peaks between the main
peaks, indicating the presence of a periodic superstructure.

\clearpage

\noindent
(a)\fig{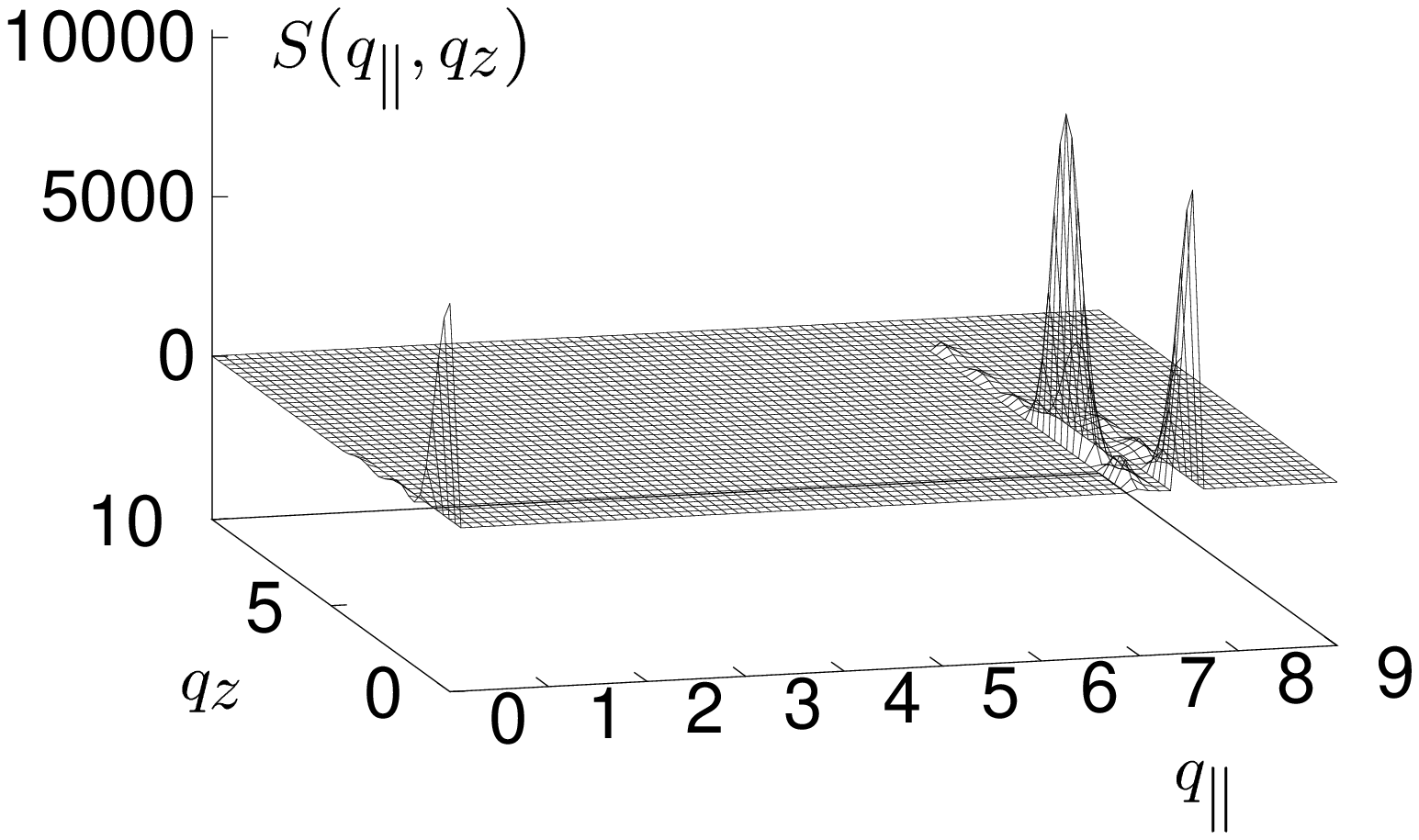}{150}{100} 
(b)\fig{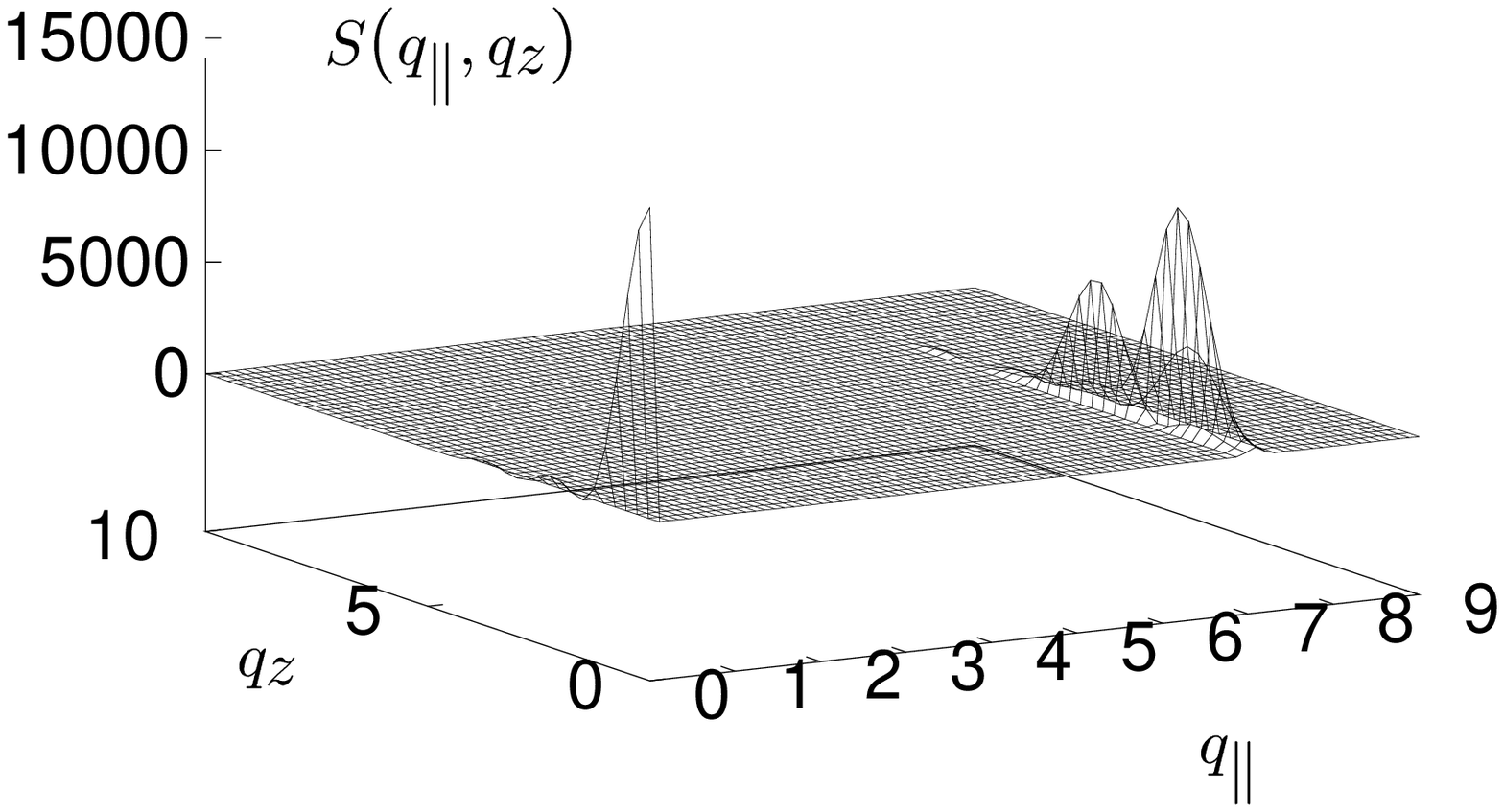}{150}{100} 

\clearpage

\noindent
Powder averaged structure factor $S(q_{\parallel},q_z)$
(a) in a state with uniform tilt towards nearest neighbors
 ($T=0.1 \epsilon/k_B, \Pi = 1 \epsilon/\sigma^2, \sigma_H = 1.1 \sigma$)
(b) in a state with uniform tilt towards next nearest neighbors
 ($T=0.1 \epsilon/k_B, \Pi = 50 \epsilon/\sigma^2, \sigma_H = 1.1 \sigma$)
(c) in a modulated state 
 ($T=0.1 \epsilon/k_B, \Pi = 1 \epsilon/\sigma^2, \sigma_H = 1.2 \sigma$).

\clearpage

\noindent
Area per molecule $A/n$ in units of $\sigma^2$ vs. temperature $T$
in units of $\epsilon/k_B$ at head size $\sigma_H=1.2 \sigma$
for different pressures $\Pi$ in units of $\epsilon/\sigma^2$ as indicated.

\clearpage

\noindent
Order parameter $R_{xy}$ in units of $\sigma^2$ vs. temperature $T$ in
units of $\epsilon/k_B$ at head size $\sigma_H=1.2 \sigma$ for different 
pressures $\Pi$ in units of $\epsilon/\sigma^2$ as indicated.

\clearpage

\noindent
Average tilt angle $\theta$ in degrees vs. temperature $T$ in
units of $\epsilon/k_B$ at head size $\sigma_H=1.2 \sigma$ for different 
pressures $\Pi$ in units of $\epsilon/\sigma^2$ as indicated.

\clearpage

\noindent
Order parameter $\Psi_6$ vs. temperature $T$ in
units of $\epsilon/k_B$ at head size $\sigma_H=1.2 \sigma$
for different pressures $\Pi$ in units of $\epsilon/\sigma^2$ as indicated.

\clearpage

\begin{figure}
\noindent
\fig{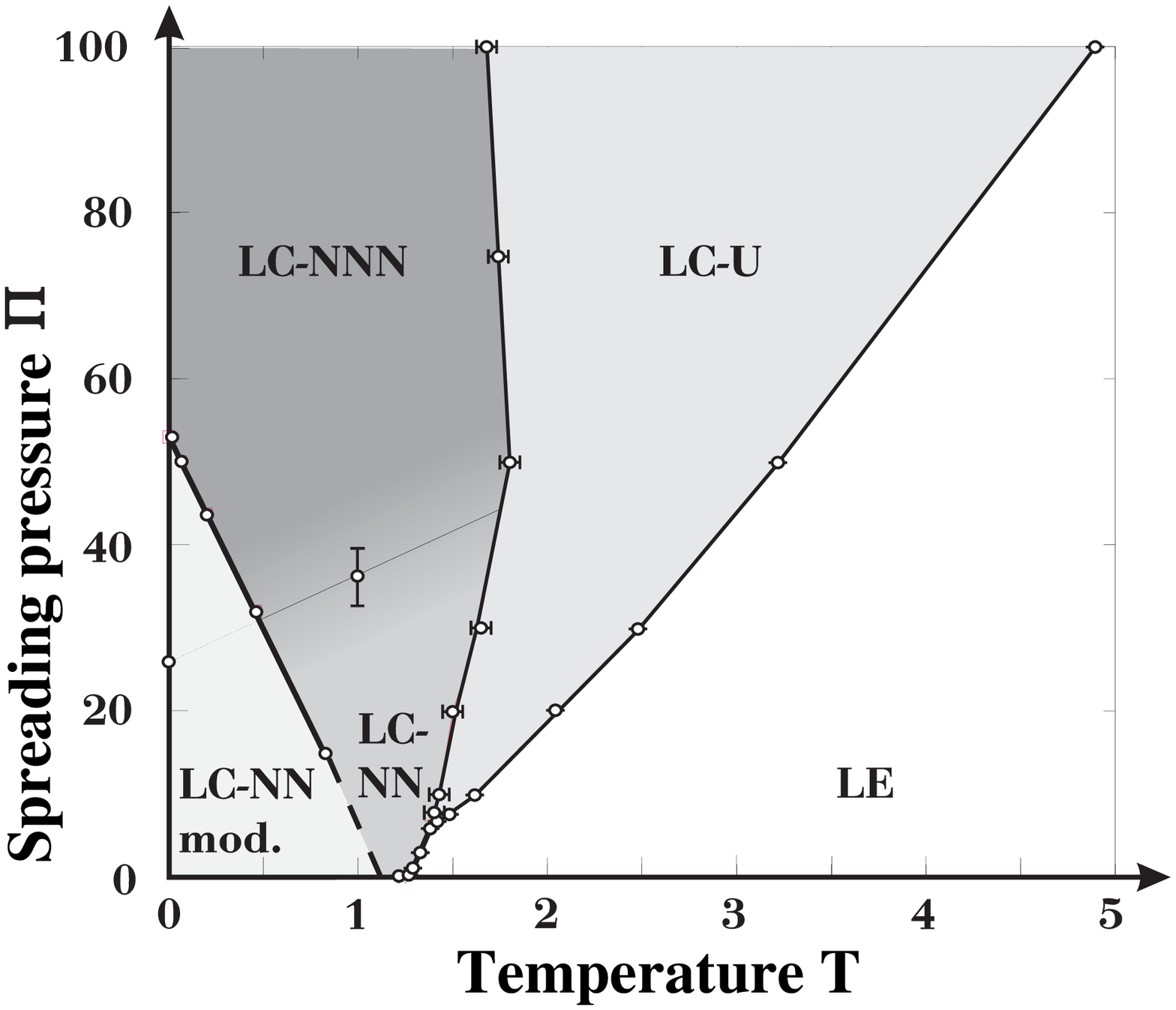}{150}{170} 
\vspace*{-2cm}
\caption{\label{fig9}}
\end{figure}
\noindent
Phase diagram of a monolayer of chains with heads of size $\sigma_H=1.2 \sigma$
and length $N=7$ beads. Units of pressure are $\epsilon/\sigma^2$, and those
of temperature $\epsilon/k_B$. LE denotes the disordered (expanded) phase, 
LC-U an untilted ordered phase, LC-NN and LC-NNN ordered phases with uniform
tilt towards nearest and next nearest neighbors, respectively, and
LC-NN mod. the modulated phase. The dotted line indicates the metastable
phase transition between two uniformly tilted phases in the region of
stability of the modulated phase. The transition between LC-NN and LC-NNN
at high temperatures is washed out and hard to localize. 

\clearpage

\begin{figure}
\noindent
\fig{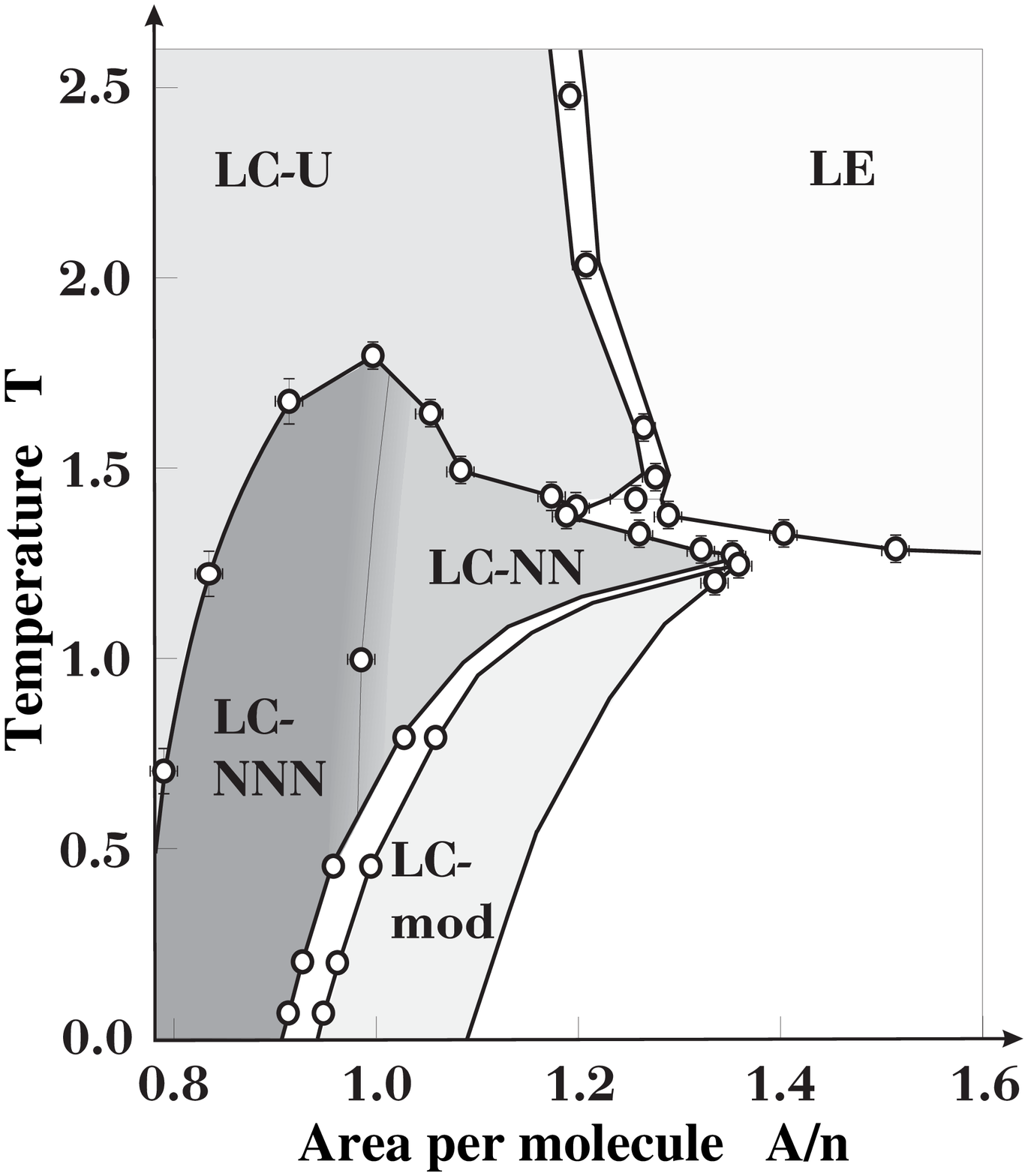}{150}{180}
\vspace*{-2cm}
\caption{\label{fig10}}
\end{figure}
\noindent
Same as Fig. \ref{fig9} in the area-temperature plane. Area per molecule
$A/n$ is given in units of $\sigma^2$.

\clearpage

\begin{figure}
\noindent
(a)\fig{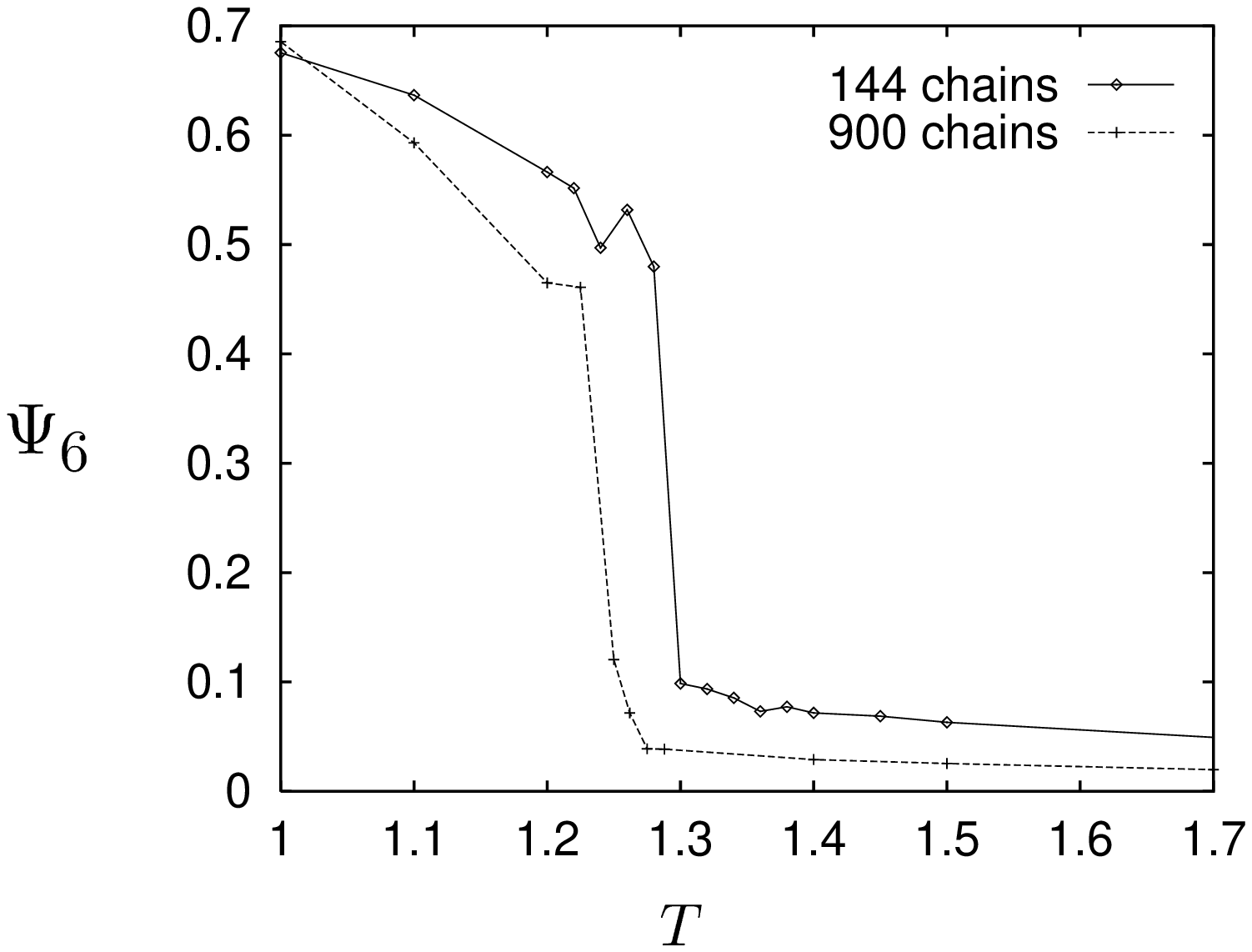}{150}{90}
(b)\fig{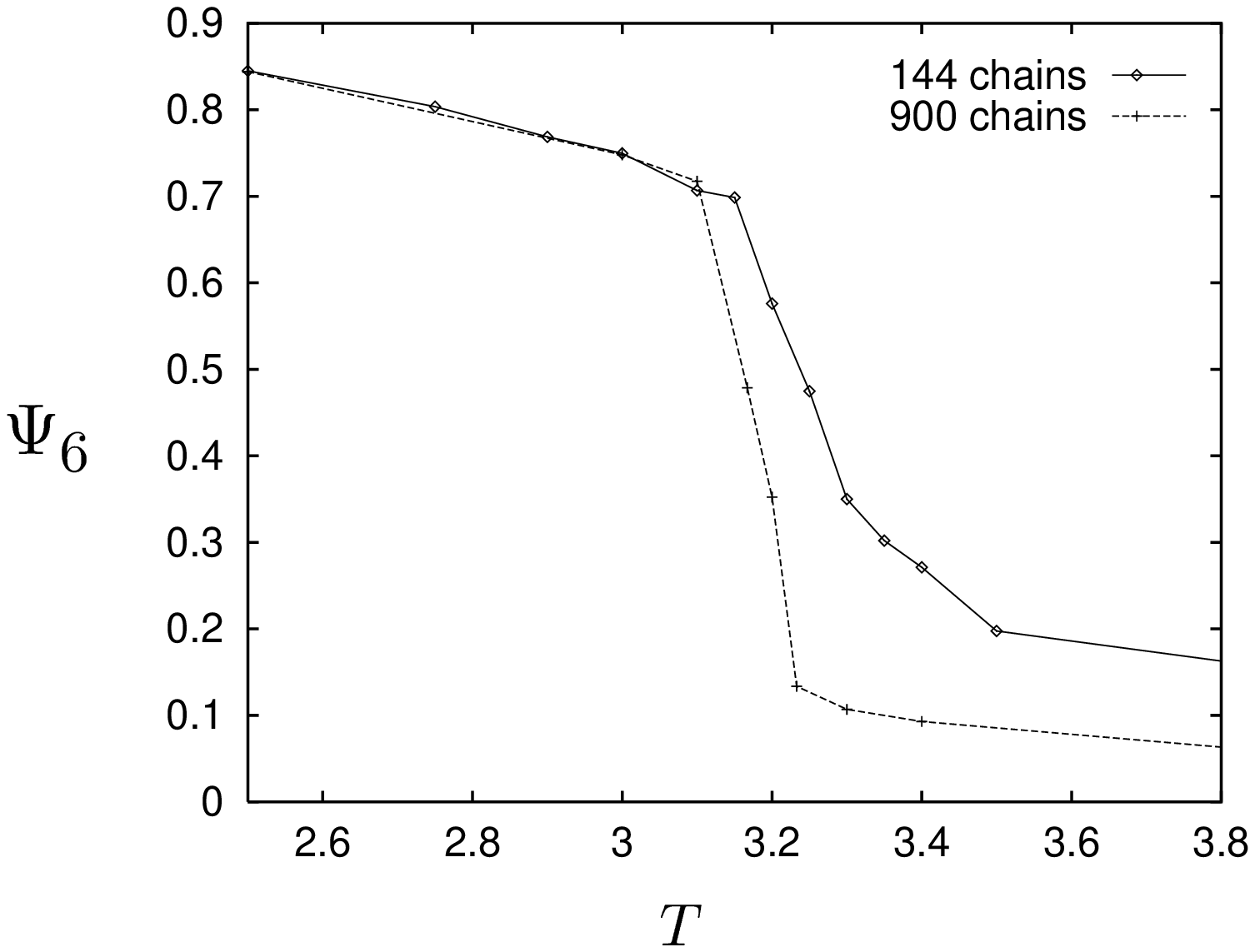}{150}{80}
\caption{\label{fig11}}
\end{figure}
\noindent
Melting order parameter $\Psi_6$ for systems of 144 chains (solid line) and
900 chains (dotted line) as a function of temperature in units of 
$\epsilon/k_B$. Pressures are $\Pi = 1 \epsilon/\sigma^2$ (a) and
$\Pi = 50 \epsilon/\sigma^2$ (b). Head size is $\sigma_H=1.2 \sigma$ and
chain length $N=7$. See text for discussion.

\clearpage

\begin{figure}
\noindent
\fig{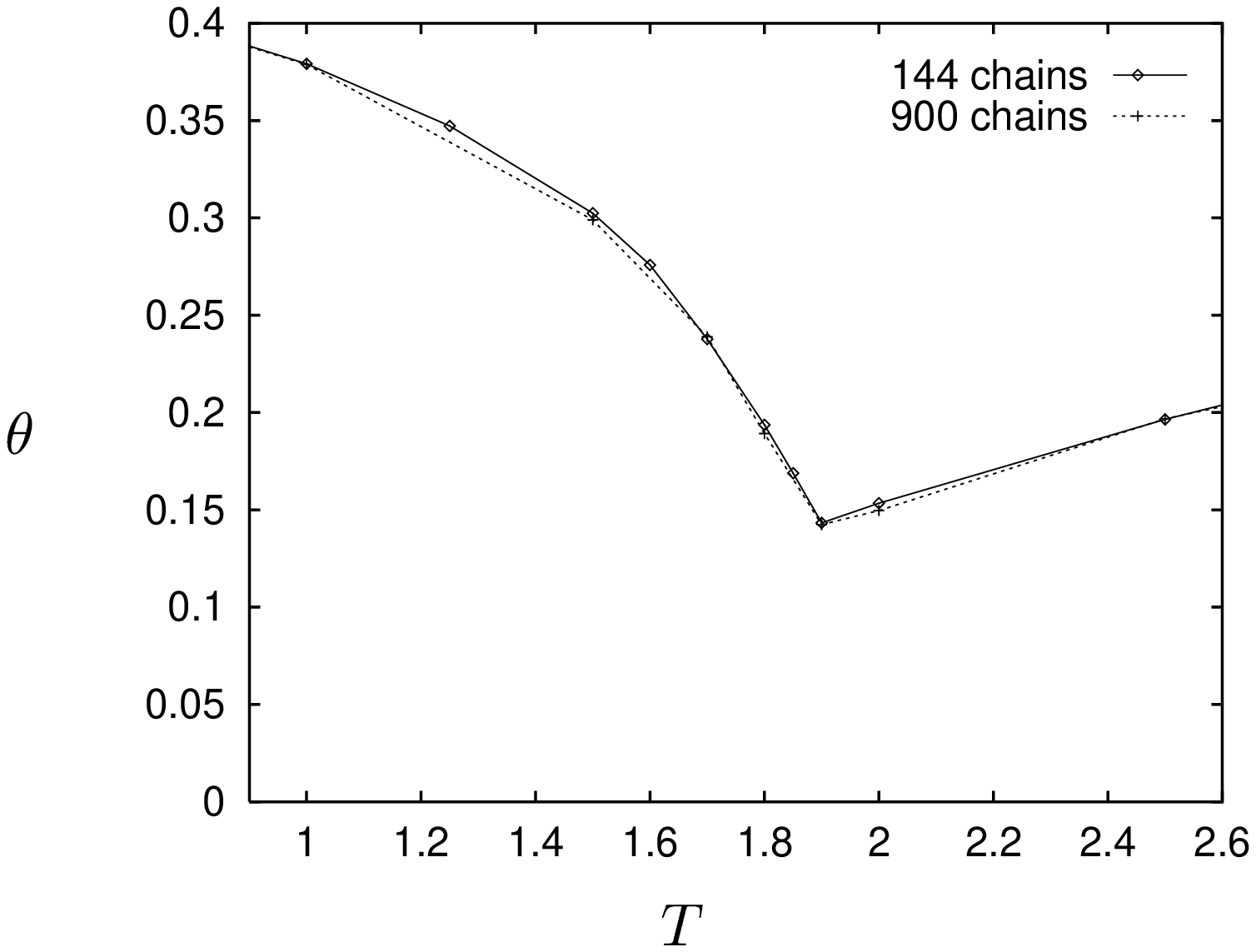}{150}{100}
\caption{\label{fig12}}
\end{figure}
\noindent
Average tilt angle $\theta$ for systems of 144 chains (solid line) and
900 chains (dotted line) vs. temperature in units of $\epsilon/k_B$.
Pressure is $\Pi = 50 \epsilon/\sigma^2$ and  
head size $\sigma_H = 1.2 \sigma$.

\clearpage

\begin{figure}
\noindent
(a)\fig{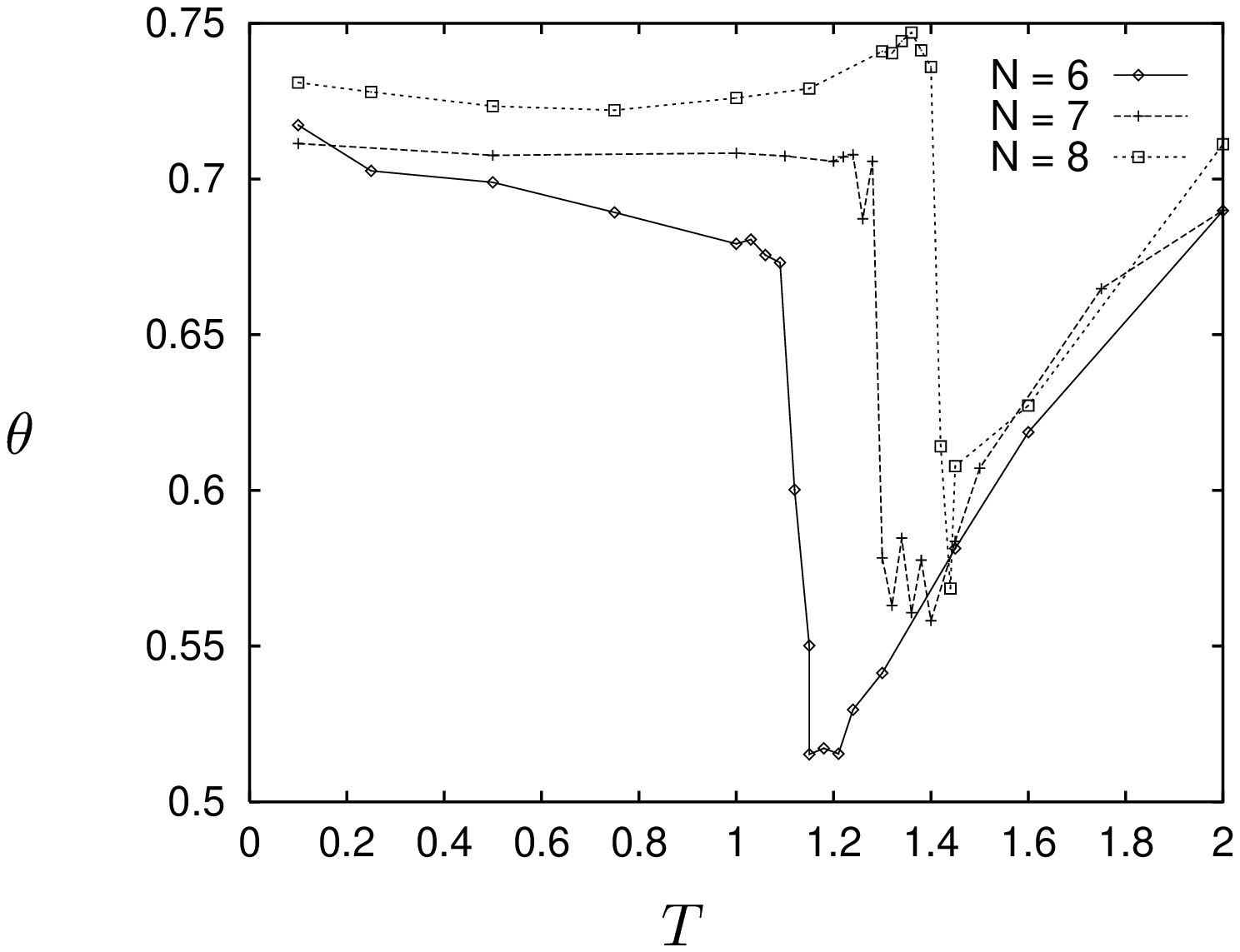}{150}{90}
(b)\fig{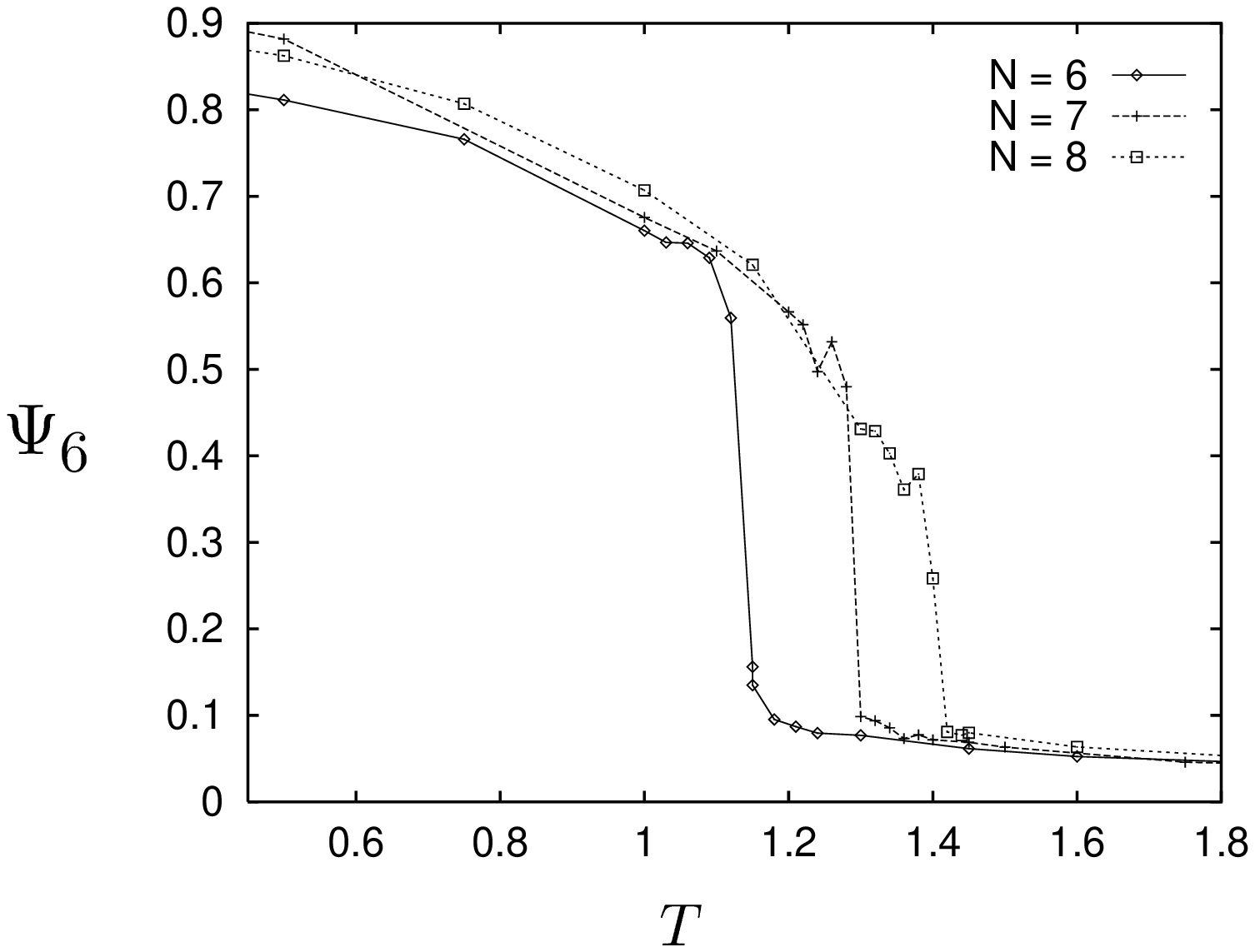}{150}{80}
\caption{\label{fig13}}
\end{figure}
\noindent
Average tilt angle $\theta$ (a) and order parameter of melting $\Psi_6$ (b)
vs. temperature (in units of $\epsilon/k_B$) at pressure 
$\Pi = 1 \epsilon/\sigma^2$ for chains of length $N=6$ (solid lines),
$N=7$ (dashed lines), and $N=8$ (dotted line).
Head size is $\sigma_H=1.2 \sigma$.

\clearpage

\begin{figure}
\noindent
(a)\fig{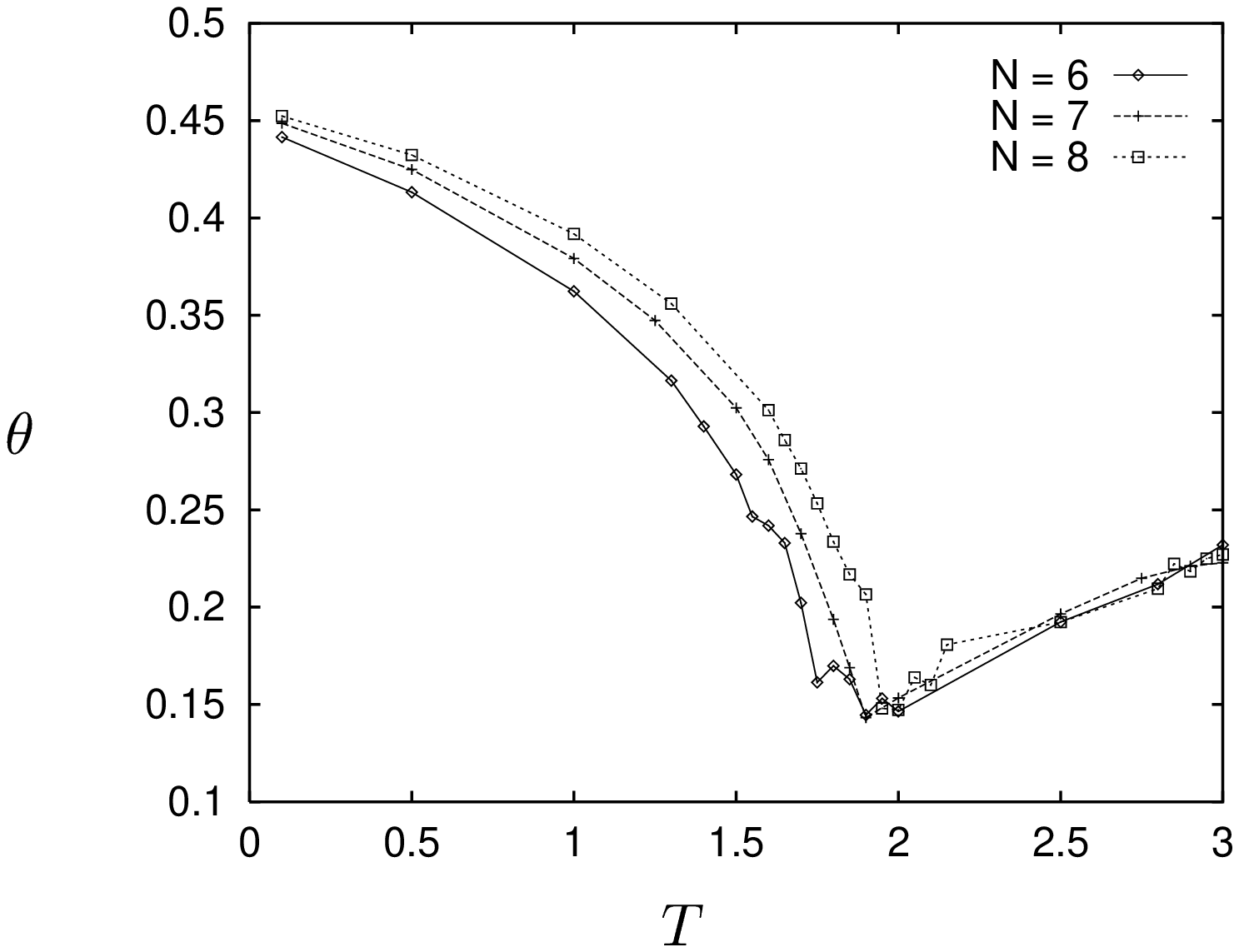}{150}{90}
(b)\fig{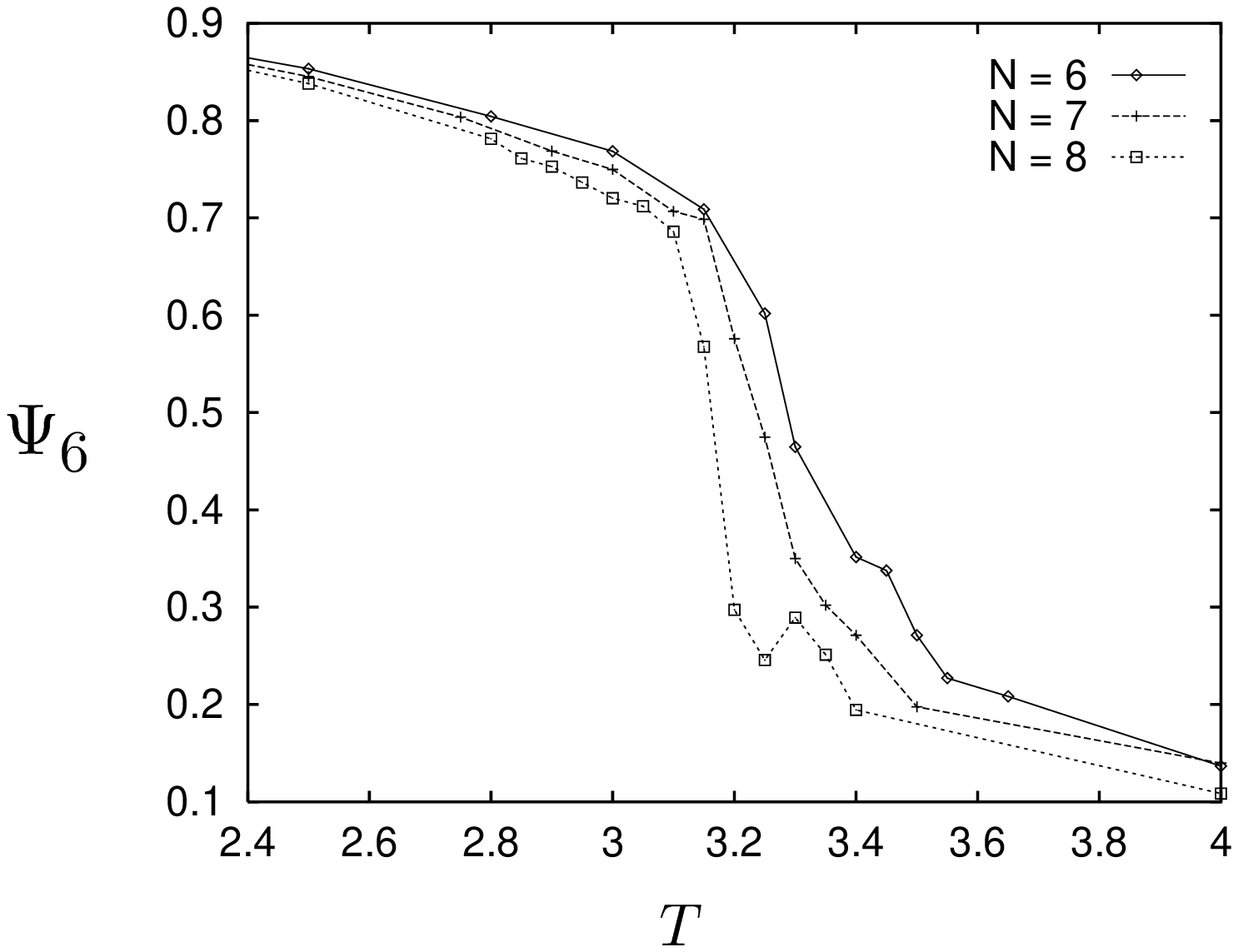}{150}{80}
\caption{\label{fig14}}
\end{figure}
\noindent
Average tilt angle $\theta$ (a) and order parameter of melting $\Psi_6$ (b)
vs. temperature (in units of $\epsilon/k_B$) at pressure 
$\Pi = 50 \epsilon/\sigma^2$ for chains of length $N=6$ (solid lines),
$N=7$ (dashed lines), and $N=8$ (dotted line).
Head size is $\sigma_H=1.2 \sigma$.

\clearpage

\begin{figure}
\noindent
(a)\fig{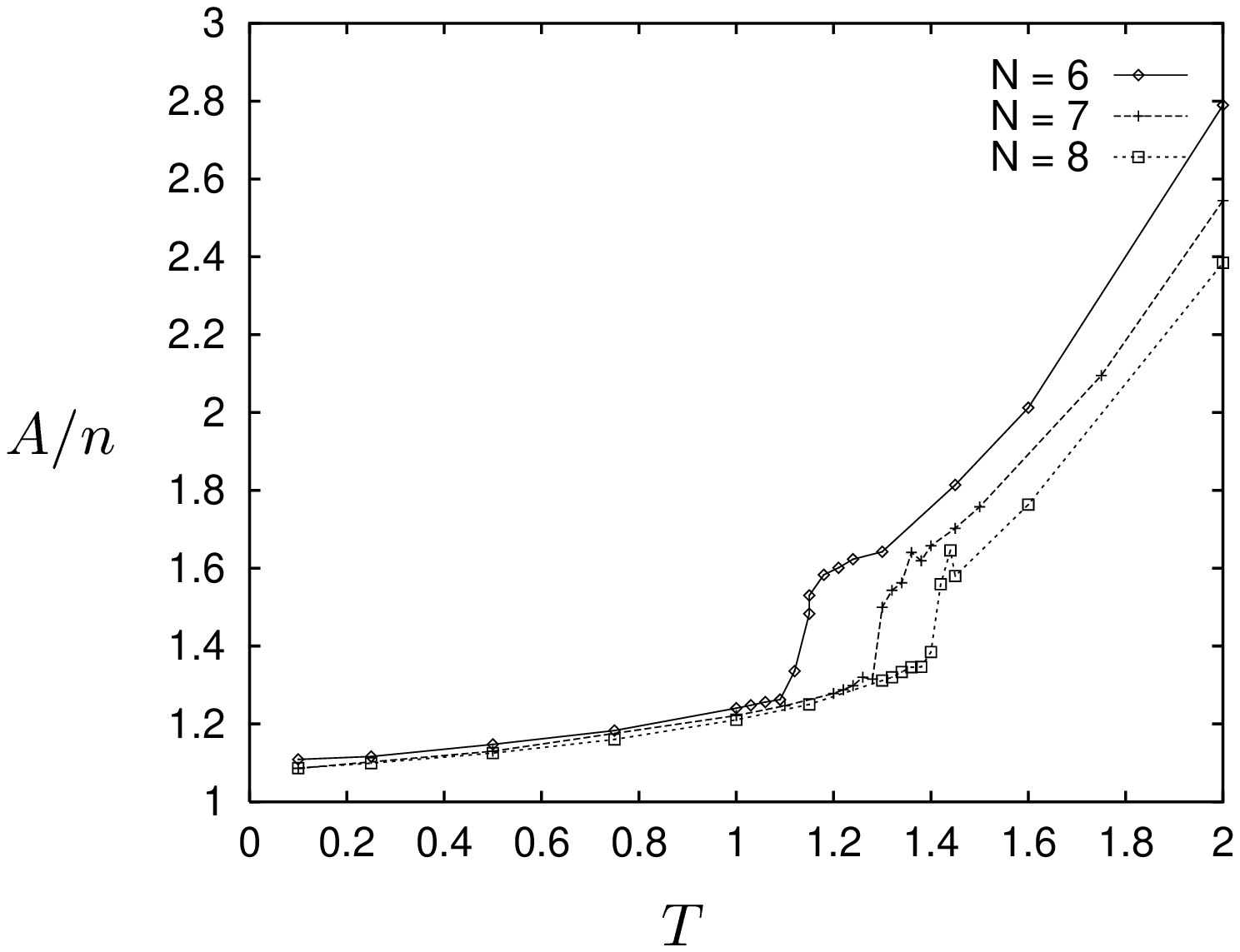}{150}{90}
(b)\fig{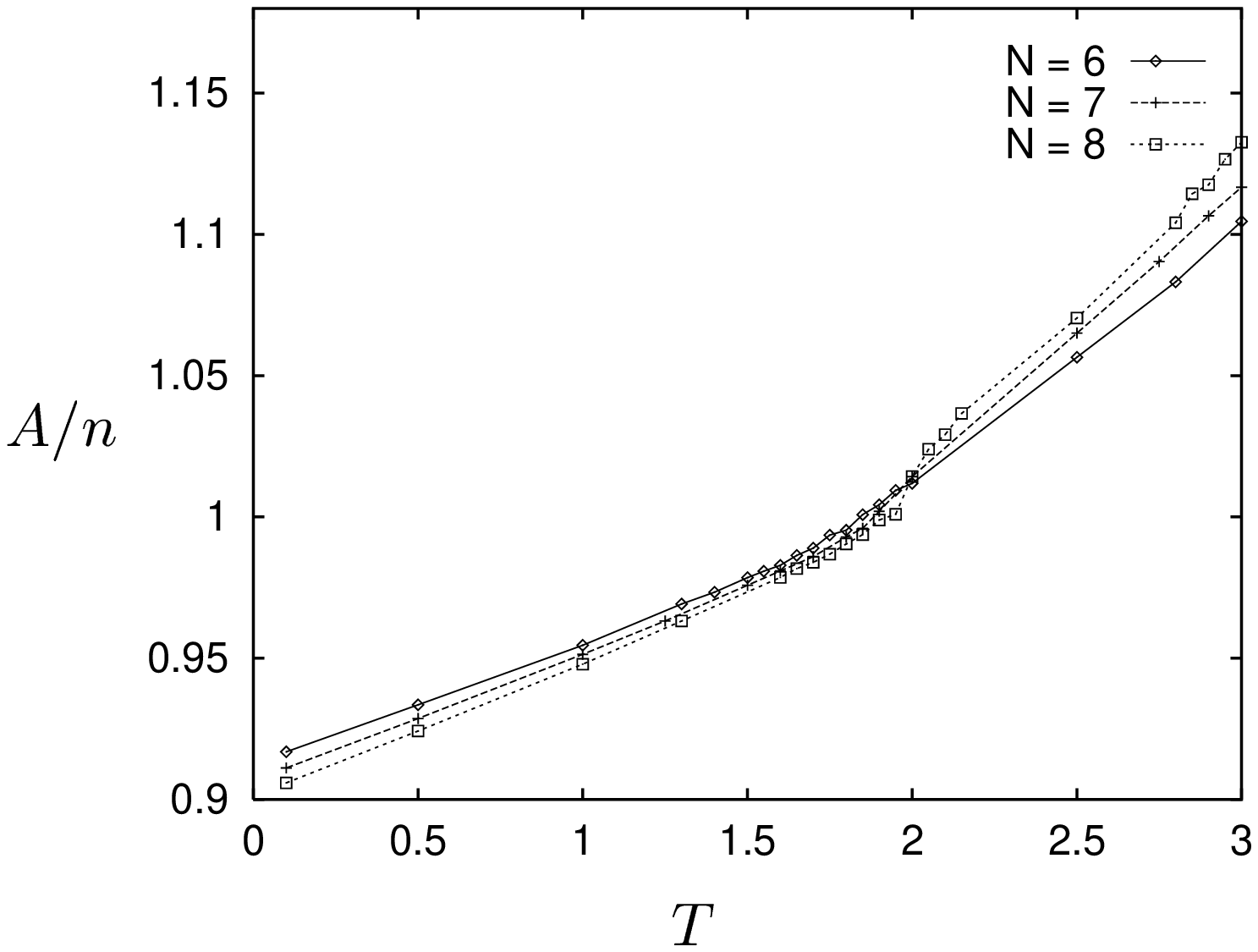}{150}{80}
\caption{\label{fig15}}
\end{figure}
\noindent
Area per molecule $A/n$ in units of $\sigma$ 
vs. temperature in units of $\epsilon/k_B$ at pressure 
$\Pi = 1 \epsilon/\sigma^2$ (a) and $\Pi = 50 \epsilon/\sigma^2$ (b) 
for chains of length $N=6$ (solid lines), $N=7$ (dashed lines), and 
$N=8$ (dotted line). Head size is $\sigma_H=1.2 \sigma$.


\begin{thebibliography}{99}
\bibitem{footnote} Strictly speaking, one has no ``Bragg'' rods
  in a strictly two-dimensional system like ours, since there is no true 
  long-range positional order. However, for all practical purposes, the
  peaks are indistinguishible from delta-singularities.
\end{thebibliography}
\end{document}